%% file: ensemble_DFPT.tex
\documentclass[citeautoscript,aps,prx,twocolumn,superscriptaddress,longbibliography]{revtex4-2}

\usepackage{filecontents}
\usepackage[colorlinks=true,
linkcolor=blue,
urlcolor=blue,
citecolor=blue]{hyperref}

\usepackage{bm}
\usepackage[utf8]{inputenc}
\usepackage{physics}
\usepackage{natbib}
\usepackage{graphicx}
\usepackage{amsmath}
\usepackage[dvipsnames]{xcolor}
\usepackage{multirow}
\usepackage{soul}
\graphicspath{{images/}}

\newcommand{\e}{\epsilon}

\begin{document}

\title{Ensemble Density-Functional Perturbation Theory: Spatial Dispersion in Metals}

\author{Asier Zabalo} 
\affiliation{Institut de Ci\`encia de Materials de Barcelona (ICMAB-CSIC), Campus UAB, 08193 Bellaterra, Spain}
\author{Massimiliano Stengel}
\affiliation{Institut de Ci\`encia de Materials de Barcelona (ICMAB-CSIC), Campus UAB, 08193 Bellaterra, Spain}
\affiliation{ICREA-Instituci\'o Catalana de Recerca i Estudis Avançats, 08010 Barcelona, Spain}

\date{\today}

\begin{abstract}
We present a first-principles methodology, within the context of linear-response theory,
that greatly facilitates the perturbative study of physical properties of
metallic crystals. 
Our approach builds on ensemble density-functional theory
[Phys. Rev. Lett. \textbf{79}, 1337 (1997)] to write the
adiabatic second-order energy as an unconstrained 
variational functional of both the wave functions and their
occupancies.
Thereby, it enables the application 
of standard tools of density-functional perturbation
theory (most notably, the ``$2n+1$'' theorem)
in metals, opening the way to an efficient and accurate calculation 
of their nonlinear and spatially dispersive responses.
We apply our methodology to phonons and strain gradients and 
demonstrate the accuracy of our implementation by computing the 
spatial dispersion coefficients of zone-center optical phonons
and the flexoelectric force-response tensor of
selected metal structures.\\
\end{abstract}

\maketitle

\section{Introduction}
In modern condensed matter physics, density-functional perturbation theory (DFPT) has emerged as the method of choice for accurately computing response properties of real materials. 
One key advantage that sets DFPT apart from alternative methods, e.g. the frozen-phonon technique \cite{PhysRevB.50.2221}, 
is its unique ability to handle incommensurate lattice distortions with arbitrary wave vectors $\mathbf{q}$ without significantly increasing the computational burden. 
Another important feature is related with the variational character of the Kohn-Sham 
energy functional: its perturbative expansion in powers of an adiabatic parameter $\lambda$
rests on the well-known $2n+1$ theorem, \cite{PhysRevA.52.1086,PhysRevA.52.1096,RevModPhys.73.515}
enabling the calculation 
of, e.g., third-order response properties with the sole knowledge of first-order wave functions. 
Interestingly, by treating the wave vector
$\mathbf{q}$ as an additional perturbation parameter,
the advantages of the $2n+1$ theorem have recently been
generalized to the long-wavelength expansion of the energy functional at an arbitrary
order in the wave vector \cite{PhysRevX.9.021050}. 
Successful application of long-wave DFPT techniques for the calculation 
of first-order spatial dispersion coefficients was demonstrated in 
several contexts,
including flexoelectric coefficients~\cite{PhysRevX.9.021050,PhysRevB.105.064101}, dynamical quadrupoles~\cite{PhysRevX.9.021050,PhysRevB.105.064101},
natural optical activity \cite{PhysRevLett.131.086902} and 
generalized Lorentz forces~\cite{PhysRevB.105.094305}.

In the context of DFPT, metals have historically been overshadowed by insulators. 
One of the main obstacles one needs to overcome when dealing with metals at zero temperature
are Brillouin Zone (BZ) sampling errors coming from the Fermi surface discontinuities.
The most effective and widely used 
strategy to deal with this issue at the ground-state level 
is the smearing technique \cite{PhysRevB.40.3616}.
In the latter approach, the sharp Fermi distribution, represented by the Heaviside step function,
is approximated by a smoother function that is a broadened approximation of the former.
The smearing approach was first
introduced in the context of DFPT by de Gironcoli \cite{PhysRevB.51.6773}, thus
enabling the computations of phonons in metals and, in turn,
of a number of thermodynamic properties that depend on phonons and electron-phonon interactions
(e.g., electrical and thermal conductivity, or superconductivity) \cite{RevModPhys.89.015003}.
In spite of its success,
the formulation proposed by de Gironcoli lacks a straightforward variational formulation.
This limitation prevents the application of the $2n+1$ theorem, which is key 
to accessing to higher-order energy derivatives.

The motivation for obtaining a theoretical framework that overcomes these obstacles
is extensive, with notable emphasis on non-linear optics (NLO) 
\cite{kaplan2023general,PhysRevX.3.021014,wu2017giant,liu2023covariant,PhysRevB.97.041101,shen1984principles}
and optical dispersion (OD) 
\cite{10.21468/SciPostPhys.14.5.118,PhysRevLett.116.077201,PhysRevB.82.245118,melrose_mcphedran_1991,
agranovich2013crystal}.
Apart from very few examples, most of the calculations that were reported 
insofar were based on semiclassical or tight-binding models. While these approaches
can provide a reliable qualitative picture in many cases, their predictive power
at the quantitative level is limited. As the experimental demand for 
reliable theoretical support is steadily growing in these areas, so is the
need for a first-principles method that is free from the aforementioned
drawbacks.
A tentative roadmap towards this ambitious goal necessarily faces some of
the long-standing technical obstacles that have slowed down progress in this
area over the years:
(i) Most first-principles attempts at calculating third-order (either nonlinear
or spatially dispersive) coefficients have
relied on cumbersome summations over a large number of unoccupied bands, with
an obvious detrimental impact on both accuracy and computational efficiency.
(ii) The effects of ``local fields'', arising due to the self-consistent 
screening of the external perturbations have seldom been accounted
for, even if their potentially dramatic impact on the response 
\cite{PhysRevLett.76.1372,PhysRevLett.131.086902} is known. 
(iii) Actual calculations are often plagued by the poorly conditioned 
nature of the dynamical response in a metal, where exceptional care in 
the computational parameters is often needed to avoid unphysical divergencies 
in the low-frequency limit.

In insulators, (i--ii) have been solved since the mid-90s, both at the 
static \cite{PhysRevB.50.5756,PhysRevB.71.125107} and dynamical \cite{PhysRevB.53.15638} level, and (iii) does not pose
any special issue as long as one works in the transparent regime.
Addressing (i--iii) in metals appears as a daunting task, as they are
all open problems in the context of the third-order response. 
As a first key step, in this work we shall focus on the issues (i--ii)
in an adiabatic (static) context, and leave the additional complications related
to the dynamical nature of the optical response to a future work.
Note that, even within the adiabatic regime, 
a plethora of outstanding physical phenomena exists that require third-order 
energy derivatives for their correct treatment, e.g., phonon anharmonicity,
nonlinear elasticity or force-response coefficients to strain gradients 
(flexocoupling coefficients, of relevance to the so-called ferroelectric 
metals~\cite{PhysRevLett.126.127601}).

To enable their first-principles calculation, here we develop a 
general perturbative framework for metallic systems by using
the ensemble density-functional theory
formalism of Marzari, Vanderbilt and Payne (MVP) \cite{PhysRevLett.79.1337}
as a conceptual basis.
The invariance of the latter with respect to unitary transformations 
within the active subspace allows us to write an \textit{unconstrained} second-order
energy functional at an arbitrary $\mathbf{q}$ vector, which is 
stationary in the first-order wave functions 
\textit{and} in the first-order occupation matrix. 
Then, mimicking the well-established procedure that
is employed with insulators \cite{PhysRevX.9.021050}, the wave vector $\mathbf{q}$ 
is treated as a perturbation parameter, which provides
(via the $2n+1$ theorem) an analytic long-wavelength 
expansion of the second-order energy functional at any desired order.
Our methodology brings the first-principles calculation of dispersion properties in metals 
to the same level of accuracy 
and efficiency as in insulators, i.e., only the knowledge of uniform field
perturbations is required to access first-order
dispersion coefficients.

This work is organized as follows. In Sec. \ref{Section_ensemble_E} we summarize
the fundamentals of ensemble density-functional theory as described in Ref. \cite{PhysRevLett.79.1337}. 
In Sec. \ref{Sec_perturbation_expansion} we perform a perturbation expansion 
of the ensemble-DFT energy functional of MVP, obtaining an unconstrained
second-order energy functional of the first-order orbitals and occupation matrices
at an arbitrary wave vector $\mathbf{q}$. 
In Sec. \ref{Sec_longwave}, following the guidelines of Ref. \cite{PhysRevX.9.021050},  
we take the first-order long-wave expansion of the aforementioned second-order energy. 
In Sec. \ref{Sec_flexocoupling}, we apply our general formalism to phonons and strain gradients, 
and validate our methodology by computing the spatial dispersion coefficients of 
zone-center phonons and the flexoelectric force-response
tensor for a number of crystals, including the well-known ferroelectric metal LiOsO$_3$.
Finally, we provide a brief summary and outlook in Sec. \ref{Sec_conclusions}. 
\section{Theory}
\subsection{Ensemble DFT}\label{Section_ensemble_E}
Here we recap the basics of ensemble DFT as formulated by
Marzari, Vanderbilt and Payne \cite{PhysRevLett.79.1337}, 
which can be regarded as a generalization of Mermin's 
formulation of finite temperature DFT \cite{PhysRev.137.A1441}.
The key assumption of MVP consists in adopting a 
matrix representation for the occupancies ($f_{ij}$) via
the following energy functional \cite{PhysRevLett.79.1337}, 
\begin{equation}\label{Eq_ensemble_DFT_E}
\begin{split}
E[\{\psi_m\},\{f_{mn} \}]=&\sum_{m,n=1}^Mf_{nm}\bra*{\psi_m}
(\hat{T}+\hat{V}_\text{ext})
\ket*{\psi_n}\\
&+E_\text{Hxc}[\rho]-\sigma S[\{ f_{mn} \}]\\
&-\sum_{m,n=1}^M\Lambda_{mn}\left( \bra*{\psi_m}\ket*{\psi_n}-\delta_{mn} \right)\\
&-\mu\left(\Tr\left(\mathbf{f}  \right)-N\right).
\end{split}
\end{equation}
Here $\hat{T}$ is the kinetic energy
operator, $\hat{V}_\text{ext}$ refers to the atomic pseudopotentials and $E_\text{Hxc}$ 
is the Hartree exchange and correlation energy, which is a functional of the
electron density,
\begin{equation}
\rho(\mathbf{r})=\sum_{m,n=1}^M f_{nm}\bra*{\psi_m}\ket*{\mathbf{r}}\bra*{\mathbf{r}}\ket*{\psi_n}.
\end{equation}
In Eq. (\ref{Eq_ensemble_DFT_E}), $\sigma$ and $S$ are, respectively, the smearing parameter and the entropy.
If the equilibrium distribution of the occupancies is chosen to follow the Fermi-Dirac (FD) 
statistics, $\sigma$ plays the role of a finite (electronic) temperature, $T$.
In practice, using a smeared distribution function is primarily aimed
at accelerating the convergence with $\mathbf{k}$-mesh density, so non-FD
forms are often preferred~\cite{PhysRevB.40.3616,PhysRevLett.82.3296}.
The Lagrange multipliers $\Lambda_{mn}$ and $\mu$ enforce the orthonormality of the wave functions
and the particle number conservation, where $N$ is the total number of electrons.
Sums are carried out for a number $M$ of \textit{active} bands with nonzero occupancies.
As long that the occupancies of the highest energy states within this \textit{active subspace} 
($\mathcal{M}$) are vanishingly small, further increasing
$M$ does not produce any change in the total energy or in any other observable
derived from Eq. (\ref{Eq_ensemble_DFT_E}).

If a diagonal representation for the occupation matrix ${\bf f}$ is enforced at all times,
Eq.~(\ref{Eq_ensemble_DFT_E}) reduces to the standard formulation~\cite{M_J_Gillan_1989,M_P_Grumbach_1994,PhysRevB.45.11372}
of Mermin's approach.
As the system evolves adiabatically in parameter space, however,
the numerical integration of the resulting electronic equations of motion 
suffers from severe ill-conditioning issues.~\cite{M_P_Grumbach_1994}
Indeed, whenever level crossings occur near the Fermi surface, the orbitals 
need to abruptly change in character (via a subspace rotation) along 
the trajectory due to the explicit imposition of the Hamiltonian gauge. 
(Sharp symmetry-protected crossings are the most catastrophic, as they imply a 
discontinuity in the adiabatic evolution of the orbitals involved.)
This is obviously not an issue in insulators, where the energy is invariant 
with respect to arbitrary unitary transformations within the occupied manifold.

The breakthrough idea of ensemble density-functional theory consists in
allowing for nonzero off-diagonal elements of the occupation matrix ($f_{ij}$),
and to treat them together with 
the wave functions ($\psi_i$) as variational parameters.
By doing so, it is easy to prove that Eq. (\ref{Eq_ensemble_DFT_E}) is 
\textit{covariant} under any unitary rotation $\mathbf{U}$ of the following type \cite{PhysRevLett.79.1337},
\begin{equation}\label{Eq_U_rotations}
\begin{split}
\quad \mathbf{f'}&=\mathbf{UfU^\dagger},\\
\ket*{\psi_m}'&=\sum_nU_{mn}^\dagger\ket*{\psi_n},
\end{split}
\end{equation}
which implies that one is no longer forced to stick to the Hamiltonian gauge.
This way, the problematic \cite{PhysRevB.60.13241,PhysRevB.62.15283} subspace rotations 
can be conveniently reabsorbed into $f_{ij}$, which 
means that the first-order variations of the wave functions are
automatically orthogonal to the active subspace, without the need of
imposing additional constraints.
\subsection{Perturbation expansion}\label{Sec_perturbation_expansion}
In the following, we assume that the system under study evolves  
from its equilibrium state, which we describe by assuming a 
dependence of the Hamiltonian on some adiabatic parameter $\lambda$. 
In the perturbative regime, we write
\begin{equation}
\hat{H}(\lambda)=\hat{H}^{(0)}+\lambda\hat{H}^{(1)}+\frac{1}{2}\lambda^2\hat{H}^{(2)}+\dots
\end{equation} 
This parametric dependence propagates in a similar way to the wave functions and the occupation matrix,
\begin{equation}
\begin{split}
\ket*{\psi_m(\lambda)}&=\ket*{\psi_m^{(0)}}+\lambda\ket*{\psi_m^{(1)}}+\dots,\\
f_{mn}(\lambda)&=f_{mn}^{(0)}+\lambda f_{mn}^{(1)}+\dots
\end{split}
\end{equation}
We now
follow the guidelines of Refs. \cite{PhysRevA.52.1096,PhysRevA.52.1086} to
recast the second-order energy as the constrained variational minimum 
of a functional that depends on both the first-order wave functions, 
$\ket*{\psi_m^{(1)}}$, and the first-order occupation matrix, $f_{mn}^{(1)}$.
(In the next section we recast this problem
as the minimization of an \textit{unconstrained} energy functional,
which poses great advantages in light of performing a long-wavelength expansion.)
We find
\begin{widetext}
	\begin{equation}\label{Eq_E_2}
	\begin{split}
	E^{(2)}_\text{con}=&2\sum_{m,n=1}^M
	\left(f_{mn}^{(0)}\bra*{\psi_n^{(1)}}\hat{H}^{(0)}\ket*{\psi_m^{(1)}}
	-\Lambda^{(0)}_{mn}\bra*{\psi_n^{(1)}}\ket*{\psi_m^{(1)}}
	\right)+2\sum_{m,n=1}^M f_{mn}^{(0)}\left(\bra*{\psi_n^{(1)}}\hat{H}^{(1)}\ket*{\psi_m^{(0)}}+\text{c.c.}\right)\\
	&+2\sum_{m,n=1}^M f_{mn}^{(1)}\bra*{\psi_n^{(0)}}\hat{H}^{(1)}\ket*{\psi_m^{(0)}}-\sigma 
	\sum_{m,n,l,k=1}^M
	 f_{mn}^{(1)}\frac{\partial^2 S}{\partial f^{(0)}_{nm}\partial f^{(0)}_{lk}}f_{lk}^{(1)}\\
	&+\int\int \rho^{(1)}(\mathbf{r})K_\text{Hxc}(\mathbf{r,r'})\rho^{(1)}(\mathbf{r'})d^3rd^3r'+\sum_{m,n=1}^M f_{mn}^{(0)} \bra*{\psi_n^{(0)}}\hat{H}^{(2)}\ket*{\psi_m^{(0)}},
	\end{split}
	\end{equation}
\end{widetext} 
where c.c. stands for complex conjugate and 
the Lagrange multipliers $\Lambda^{(0)}_{mn}$ are related to the matrix elements of the ground state Hamiltonian,
$\Lambda^{(0)}_{mn}=f^{(0)}_{nm}\bra*{\psi_m^{(0)}}\hat{H}^{(0)}\ket*{\psi_n^{(0)}}$, with
$f^{(0)}_{nm}=f_m\delta_{nm}$.
(Note that the energy functional $E^{(2)}$ is defined as the second derivative
of the total energy; this convention differs by a factor of two with respect to
Refs. \cite{PhysRevX.9.021050,PhysRevB.55.10337,PhysRevA.52.1086,PhysRevA.52.1096}.)
The subscript ``con" in Eq. (\ref{Eq_E_2}) indicates that this energy functional is minimized subject
to certain constraints. In this case, we impose that the ground-state wave functions are orthogonal to 
the first-order wave functions,
\begin{equation}
\bra*{\psi_m^{(0)}}\ket*{\psi_n^{(1)}}=0, \quad\forall m,n\in \mathcal{M},
\end{equation}
which is also known as the \textit{parallel transport gauge} \cite{PhysRevA.52.1096}.
We shall explain all the new terms appearing in Eq. (\ref{Eq_E_2}) in the following. 
The second derivative of the 
entropy with respect to the occupation matrix appears in the second line of 
Eq. (\ref{Eq_E_2}). By assuming a diagonal representation for the unperturbed occupation
matrix, one can show that the following holds,
\begin{equation}
\sigma \frac{\partial^2 S}{\partial f^{(0)}_{mn}\partial
f^{(0)}_{kl}}=  \frac{ \delta_{mk}\delta_{nl} }{\bar{f}_{mn}},
\end{equation}
where the matrix $\bar{f}_{mn}\equiv G(\epsilon_m,\epsilon_n)$ is defined as \cite{PhysRevB.62.15283}
\begin{equation}\label{Eq_G}
G(x,y)=
\begin{cases}
\frac{f(x)-f(y)}{x-y}&,\quad\text{if } x\neq y,\\[8pt]
\frac{1}{2}\left(\frac{\partial f(x)}{\partial x}
+ \frac{\partial f(y)}{\partial y}\right)&, \quad\text{if } x=y.
\end{cases}
\end{equation}
(In the numerical implementation we set a finite tolerance to test the equality of
$x$ and $y$, hence the need for the symmetrization in the second line; similar considerations
apply to Eq.~(\ref{Eq_calligraphic_F}) later on.)
The third line in 
Eq. (\ref{Eq_E_2}) contains the Hartree exchange and correlation (Hxc) kernel,
\begin{equation}
K_\text{Hxc}(\mathbf{r,r'})=\frac{\delta V_\text{Hxc}(\mathbf{r})}{\delta \rho(\mathbf{r'})}=
\frac{\delta^2 E_\text{Hxc}}{\delta \rho(\mathbf{r})\delta \rho(\mathbf{r'})},
\end{equation}
and $\rho^{(1)}(\mathbf{r})$ is the first-order electron density, which depends both on the 
first-order wave functions as well as on the first-order occupation matrix, 
\begin{equation}
\rho^{(1)}=\frac{\partial\rho}{\partial\psi}\psi^{(1)}+\frac{\partial \rho}{\partial f}f^{(1)}.
\end{equation}
The stationary condition on the first-order occupation matrix allows us to find a solution for $f^{(1)}_{mn}$,
\begin{equation}\label{Eq_f1}
\frac{\delta E^{(2)}}{\delta f_{mn}^{(1)}}=0 \longrightarrow 
f_{mn}^{(1)}=\bar{f}_{mn}\bra*{\psi_m^{(0)}}\mathcal{\hat{H}}^{(1)}\ket*{\psi_n^{(0)}},
\end{equation}
where the calligraphic Hamiltonian indicates that SCF terms have been included, 
$\mathcal{\hat{H}}^{(1)}=\hat{H}^{(1)}+\hat{V}^{(1)}$, with
\begin{equation}
V^{(1)}(\mathbf{r})=\int  K_\text{Hxc}(\mathbf{r,r'})\rho^{(1)}(\mathbf{r'})d^3r'. 
\end{equation}
\subsection{Unconstrained variational formulation}

At this stage, in complete analogy with the insulating case \cite{PhysRevX.9.021050} and with the forthcoming goal of performing a long-wavelength expansion of the energy functional, we can write an
\textit{unconstrained variational functional} for $E^{(2)}$ as follows, 
\begin{widetext}
	\begin{equation}\label{Eq_E2_unconstrained}
	\begin{split}
	E^{(2)}=&2\sum_{m=1}^M f_m\bra*{\psi_m^{(1)}}(\hat{H}^{(0)}+a\hat{P}-\epsilon_m^{(0)})\ket*{\psi_m^{(1)}}
	+2\sum_{m=1}^M f_m\left(\bra*{\psi_m^{(1)}}\hat{Q}\hat{H}^{(1)}\ket*{\psi_m^{(0)}}+\text{c.c.}\right)\\
	&+2\sum_{m=1}^M f^{(1)}_{mn}\bra*{\psi_n^{(0)}}\hat{H}^{(1)}\ket*{\psi_m^{(0)}}
	-\sum_{m,n=1}^M f_{mn}^{(1)}\frac{\epsilon^{(0)}_n-\epsilon^{(0)}_m}{f_n-f_m}f^{(1)}_{nm}\\
	&+\int\int \rho^{(1)}(\mathbf{r})K_\text{Hxc}(\mathbf{r,r'})\rho^{(1)}(\mathbf{r'})d^3rd^3r'
	+\sum_{m}^M f_m\bra*{\psi_m^{(0)}}\hat{H}^{(2)}\ket*{\psi_m^{(0)}},
	\end{split}
	\end{equation}
\end{widetext}
where $a$ is a parameter with the dimension of energy that ensures the stability 
of the functional \cite{PhysRevX.9.021050,RevModPhys.73.515} and the operators
\begin{equation}
\begin{split}
\hat{P}=\sum_{m=1}^M\ket*{\psi_m^{(0)}}\bra*{\psi_m^{(0)}},\quad \hat{Q}=1-\hat{P},
\end{split}
\end{equation}
are projectors onto and out of the active subspace, respectively. These also are also relevant in the 
first-order electron density,
\begin{equation}\label{Eq_n1}
\begin{split}
\rho^{(1)}(\mathbf{r})=&\sum_{m=1}^M f_m 
\bra*{\psi_m^{(1)}}\hat{Q}\ket*{\mathbf{r}}\bra*{\mathbf{r}}\ket*{\psi_m^{(0)}}
+\text{c.c.}\\
&+\sum_{m,n=1}^Mf^{(1)}_{mn}\bra*{\psi_n^{(0)}}\ket*{\mathbf{r}}\bra*{\mathbf{r}}\ket*{\psi_m^{(0)}}.
\end{split}
\end{equation}
(Note that, unlike in the
insulating case, $\hat{P}$ does not correspond to the ground-state density operator.)
The stationary condition on the first-order wave functions, 
$\delta E^{(2)}/\delta\bra*{\psi_m^{(1)}}=0$, leads to a standard Sternheimer equation, as proposed by
Baroni \textit{et al.} \cite{RevModPhys.73.515},
\begin{equation}\label{Eq_Sternheimer}
(\hat{H}^{(0)}+a\hat{P}-\epsilon_m^{(0)})\ket*{\psi_m^{(1)}}=-\hat{Q}\hat{\mathcal{H}}^{(1)}\ket*{\psi_m^{(0)}}.
\end{equation}

\subsection{Nonstationary formulas}
Plugging the stationary conditions Eq. (\ref{Eq_f1}) and  Eq. (\ref{Eq_Sternheimer})
into Eq. (\ref{Eq_E2_unconstrained}) results in the following
\textit{nonstationary} (nonst) expression for the second-order energy,
\begin{equation}\label{Eq_E2_nonst}
\begin{split}
E^{(2)}_\text{nonst}=&\sum_{m=1}^M f_m
\bra*{\psi_m^{(1)}}\hat{H}^{(1)}\ket*{\psi_m^{(0)}}+\text{c.c.}\\
&+\sum_{m,n=1}^M f_{mn}^{(1)}\bra*{\psi_n^{(0)}}\hat{H}^{(1)}\ket*{\psi_m^{(0)}}\\
&+\sum_{m=1}^M f_m \bra*{\psi_m^{(0)}}\hat{H}^{(2)}\ket*{\psi_m^{(0)}}.
\end{split}
\end{equation}
Interestingly, the only difference between Eq. (\ref{Eq_n1}) and Eq. (\ref{Eq_E2_nonst}) is that
the first-order \textit{external perturbation}, $\hat{H}^{(1)}$, appearing in the second-order energy is substituted with
the operator $\ket*{\mathbf{r}}\bra*{\mathbf{r}}$ in the first-order electron density expression.
We can even achieve a more compact version of the above by writing both the second-order energy
and the first-order electron density as a trace,
\begin{equation}
E^{(2)}_\text{nonst}=\Tr \left(
\hat{\rho}^{(1)} \hat{H}^{(1)} 
+ \hat{\rho}^{(0)} \hat{H}^{(2)} 
\right)
\end{equation}
and
\begin{equation}
\rho^{(1)}=\Tr\left(
\hat{\rho}^{(1)}\ket*{\mathbf{r}}\bra*{\mathbf{r}}\right),
\end{equation}
where the ground-state density operator is given by
\begin{equation}
\hat{\rho}^{(0)}=\sum_{m=1}^M \ket*{\psi_m^{(0)}}f_m\bra*{\psi_m^{(0)}},
\end{equation}
and we have defined the \textit{first-order density operator} as 
\begin{equation}\label{Eq_rho1}
\begin{split}
\hat{\rho}^{(1)}=&\sum_{m=1}^M  \ket*{\psi_m^{(0)}}f_m\bra*{{\psi}_m^{(1)}} +
\text{c.c}\\
&+\sum_{m,n=1}^M \ket*{\psi_m^{(0)}}f^{(1)}_{mn}\bra*{\psi_n^{(0)}}.
\end{split}
\end{equation}
\subsection{Relation to De Gironcoli's approach}
Our formalism naturally recovers De Gironcoli's standard expressions for metals.
We start by defining the following \textit{tilded} first-order wave functions,
\begin{equation}\label{Eq_tilde_1wf}
\begin{split}
\ket*{\tilde{\psi}_m^{(1)}}=\ket*{\psi_m^{(1)}}+\frac{1}{2f_m}
\sum_{n=1}^M \ket*{\psi_n^{(0)}}f^{(1)}_{mn}.
\end{split}
\end{equation}
The first-order density matrix is then given by
\begin{equation}
\hat{\rho}^{(1)}=\sum_{m=1}^M \ket*{\psi_m^{(0)}}f_m \bra*{\tilde{\psi}_m^{(1)}}
+\text{c.c.},
\end{equation}
which, in turn, exactly reduces to our Eq. (\ref{Eq_rho1}).
The nonstationary second-order energy and the first-order electron density can then be expressed succinctly using this notation as
\begin{equation}\label{Eq_E2_nonst_tilde}
\begin{split}
E^{(2)}_\text{nonst}=&\sum_{m=1}^M f_m \bra*{\tilde{\psi}_m^{(1)}}\hat{H}^{(1)}
\ket*{\psi_m^{(0)}}+\text{c.c.}\\
&+\sum_{m=1}^M f_m \bra*{\psi_m^{(0)}}\hat{H}^{(2)}\ket*{\psi_m^{(0)}}
\end{split}
\end{equation}
and
\begin{equation}\label{Eq_rho1_nonst_tilde}
\rho^{(1)}=\sum_{m=1}^M f_m \bra*{\tilde{\psi}_m^{(1)}}\ket*{\mathbf{r}}\bra*{\mathbf{r}}
\ket*{\psi_m^{(0)}}+\text{c.c.},
\end{equation}
where this expression for the first-order electron density, Eq. (\ref{Eq_rho1_nonst_tilde}), coincides with Eq. (10) of 
Ref. \cite{PhysRevB.51.6773}. Our tilded first-order wave functions defined here can
therefore be regarded as the $\Delta \phi_i(\mathbf{r})$ wave functions of de Gironcoli's 
approach in Ref \cite{PhysRevB.51.6773}.
Interestingly, Eq. (\ref{Eq_E2_nonst_tilde}) and Eq. (\ref{Eq_rho1_nonst_tilde}) resemble the
expressions that are commonly used in insulators. Here, however, the \textit{tilded} first-order
wave functions take into account the subspace unitary rotations in the active subspace, via the second
term on the right-hand side of Eq. (\ref{Eq_tilde_1wf}); this term is absent in insulators.

\subsection{Parametric derivative of operators}\label{Sec_deriv_operators}
Eq.~(\ref{Eq_rho1}) can be regarded as special case of a more general 
rule for differentiating operators along adiabatic paths in parameter space.
We establish this rule in the following, since it is key to 
the long-wave expansion of the second-order energy functional that
we perform in the next section.

Consider an operator  
in the following form,
\begin{equation}
\hat{O} = \sum_{m=1}^M |\psi_m^{(0)}\rangle h(\e^{(0)}_m) \langle \psi_m^{(0)}|,
\end{equation}
where $h(\e^{(0)}_m)$ is a real and differentiable function of the eigenvalue $\e^{(0)}_m$. 
The derivative of $\hat{O}$ with respect to an adiabatic parameter $\lambda$ is then given by
\begin{equation}\label{diff_rule}
\begin{split}
\frac{\partial \hat{O}}{\partial \lambda} =& \sum_{m=1}^M  \ket*{\psi_m^{(0)}}h(\e^{(0)}_m)\bra*{{\psi}_m^{\lambda}} +
\text{c.c.}\\
 & + \sum_{m,l=1}^M \mathcal{G}(\e^{(0)}_l,\e^{(0)}_m) |\psi_m^{(0)} \rangle  \langle \psi_m^{(0)} | \mathcal{H}^\lambda |\psi_l^{(0)} \rangle  \langle \psi_l^{(0)}|,
\end{split}
\end{equation}
where $\mathcal{G}$ is defined as in our Eq. (\ref{Eq_G}), only replacing $f$ with $h$ therein.
This result essentially corresponds to Eq. (20) of Ref. \cite{liu2023covariant}, but recast in an ensemble-DFPT form. 
Its proof rests on the following two rules,
\begin{subequations}\label{Eq_parametric_deriv_rules}
\begin{align}
\frac{\partial h(\e^{(0)}_m)}{\partial\lambda}=&\frac{\partial h(\e^{(0)}_m)}{\partial\e^{(0)}_m}
\bra*{\psi_m^{(0)}}\hat{\mathcal{H}}^\lambda\ket*{\psi_m^{(0)}}, \\
\frac{\partial \ket*{\psi_m^{(0)}}}{\partial\lambda}=&\ket*{\psi^\lambda_m}+
\sum_{\substack{n=1 \\ n\neq m}}^M
\frac{\bra*{\psi_n^{(0)}}\hat{\mathcal{H}}^\lambda\ket*{\psi_m^{(0)}}}{\epsilon_m^{(0)}-\epsilon_n^{(0)}},
\end{align}
\end{subequations}
where the term $n=m$ is not included in the summation.
Note that, when applied separately, Eq.~(\ref{Eq_parametric_deriv_rules}) require that there be no 
degeneracies in the spectrum; conversely, Eq.~(\ref{diff_rule}) is valid in the general case.
Whenever $h$ corresponds to the occupation function $f$, the operator $\hat{O}$
reduces to the ground-state density operator and Eq.~(\ref{diff_rule}) becomes
Eq.~(\ref{Eq_rho1}).

\section{Long-wave expansion of the second-order energy functional}\label{Sec_longwave}
\subsection{Monochromatic perturbations}
We now apply the formalism presented in the previous subsection to the case of 
a monochromatic perturbation, modulated at a wave vector $\mathbf{q}$.
By exploiting linearity, the first-order Hamiltonian is 
conveniently written as a phase times a cell-periodic part 
\cite{PhysRevB.55.10337,PhysRevX.9.021050}, such that
\begin{equation}
\hat{H}^{(1)}(\mathbf{r,r'})=e^{i\mathbf{q\cdot r}}\hat{H}_\mathbf{q}^{(1)}(\mathbf{r,r'}),
\end{equation}
and the Kohn-Sham wave functions are expressed as 
\begin{equation}
\psi_{m\mathbf{k}}^{(0)}(\mathbf{r})=e^{i\mathbf{k\cdot r}}u_{m\mathbf{k}}^{(0)}(\mathbf{r}),
\end{equation}
where $u_{m\mathbf{k}}^{(0)}(\mathbf{r})$ are the cell-periodic Bloch functions. We can now 
write the second-order energy functional as a stationary (st) functional (with respect to the wave functions and the occupation matrix) plus a nonvariational (nv) contribution,
\begin{equation}\label{Eq_E2_st_plus_nv}
E_\mathbf{q}^{\lambda_1\lambda_2}=E^{\lambda_1\lambda_2}_{\text{st},\mathbf{q}}+
E^{\lambda_1\lambda_2}_{\text{nv},\mathbf{q}}.
\end{equation}
For the sake of generality, we consider mixed derivatives with respect to two arbitrary perturbations, 
$\lambda_1$ and $\lambda_2$. 
(Applications to the specific cases of phonons and 
strains are discussed in Section~\ref{Sec_flexocoupling}.)
The stationary part can be written as follows, 
\begin{equation}\label{Eq_E2_q}
\begin{split}
E^{\lambda_1\lambda_2}_{\text{st},\mathbf{q}}=&\int_\text{BZ}[d^3k]\left(
\bar{E}_\mathbf{k,q}^{\lambda_1\lambda_2}+\bar{E}_\mathbf{k+q,-q}^{\lambda_1\lambda_2 \,*}
+\Delta E_\mathbf{k,q}^{\lambda_1\lambda_2}
\right)\\
&+\int_\Omega\int \rho^{\lambda_1*}_\mathbf{q}(\mathbf{r})K_\mathbf{q}(\mathbf{r,r'})
\rho^{\lambda_2}_\mathbf{q}(\mathbf{r'}) 
d^3rd^3r' ,
\end{split}
\end{equation}
where the shorthand notation $[d^3k]=\Omega/(2\pi)^3 d^3k$ is used for the Brillouin Zone (BZ)
integration and 
\begin{equation}
K_\mathbf{q}(\mathbf{r,r'})=K_\text{Hxc}(\mathbf{r,r'})e^{i\mathbf{q}\cdot(\mathbf{r'-r})}
\end{equation}
is the phase-corrected Coulomb and exchange-correlation kernel.
The second line of Eq. (\ref{Eq_E2_q}) explicitly depends on the first-order electron densities,
\begin{equation}\label{Eq_rho_1_q}
\begin{split}
\rho_\mathbf{q}^\lambda(\mathbf{r})=\int_\text{BZ}[d^3k]\Bigg[
&\sum_{m=1}^M \Big(f_{m\mathbf{k}}
\bra*{u_{m\mathbf{k}}^{(0)}}\ket*{\mathbf{r}}\bra*{\mathbf{r}}\hat{Q}_\mathbf{k+q}
\ket*{u_{m\mathbf{k,q}}^\lambda}\\
&+ f_{m\mathbf{k+q}}\bra*{u_{m\mathbf{k+q,-q}}^\lambda}\hat{Q}_\mathbf{k}\ket*{\mathbf{r}}
\bra*{\mathbf{r}}\ket*{u_{m\mathbf{k+q}}^{(0)}}
\Big)\\
&+\sum_{m,n=1}^M
\bra*{u_{m\mathbf{k}}^{(0)}}\ket*{\mathbf{r}}\bra*{\mathbf{r}}\ket*{u_{n\mathbf{k+q}}^{(0)}}
f_{n\mathbf{k+q},m\mathbf{k}}^\lambda
\Bigg],
\end{split}
\end{equation}
where $\ket*{u_{m\mathbf{k,q}}^\lambda}$ and $f_{n\mathbf{k+q},m\mathbf{k}}^\lambda$
are the first-order trial wave functions and occupation matrices, respectively.
We shall name the new symbols appearing in Eq. (\ref{Eq_E2_q}) as the \textit{wave function}
($\bar{E}$) and \textit{occupation} 
($\Delta E$) contributions, which are given by
\begin{equation}
\begin{split}
\bar{E}^{\lambda_1\lambda_2}_\mathbf{k,q}=&\sum_{m=1}^M f_{m\mathbf{k}}
\bra*{u_{m\mathbf{k,q}}^{\lambda_1}}
(\hat{H}^{(0)}_\mathbf{k+q}+a\hat{P}_\mathbf{k+q}-\epsilon_{m\mathbf{k}}^{(0)})
\ket*{u_{m\mathbf{k,q}}^{\lambda_2}}\\
&+\sum_{m=1}^M f_{m\mathbf{k}}\bra*{u_{m\mathbf{k,q}}^{\lambda_1}}\hat{Q}_\mathbf{k+q}
\hat{H}^{\lambda_2}_\mathbf{k,q}\ket*{u_{m\mathbf{k}}^{(0)}}\\
&+\sum_{m=1}^M f_{m\mathbf{k}}\bra*{u_{m\mathbf{k}}^{(0)}}(\hat{H}^{\lambda_1}_\mathbf{k,q})^\dagger
\hat{Q}_\mathbf{k+q}\ket*{u_{m\mathbf{k,q}}^{\lambda_2}},
\end{split}
\end{equation}
and
\begin{equation}\label{Eq_Delta_E2}
\begin{split}
\Delta E^{\lambda_1\lambda_2}_\mathbf{k,q}=&
\sum_{m,n=1}^Mf^{\lambda_1}_{m\mathbf{k},n\mathbf{k+q}}
\bra*{u_{n\mathbf{k+q}}^{(0)}}\hat{H}^{\lambda_2}_\mathbf{k,q}\ket*{u_{m\mathbf{k}}^{(0)}}\\
&+\sum_{m,n=1}^M \bra*{u_{m\mathbf{k}}^{(0)}}(\hat{H}^{\lambda_1}_\mathbf{k,q})^\dagger
\ket*{u_{n\mathbf{k+q}}^{(0)}}
f_{n\mathbf{k+q},m\mathbf{k}}^{\lambda_2}\\
&-\sum_{m,n=1}^M f^{\lambda_1}_{m\mathbf{k},n\mathbf{k+q}}
\frac{\epsilon^{(0)}_{n\mathbf{k+q}}-\epsilon^{(0)}_{m\mathbf{k}}}
{f_{n\mathbf{k+q}}-f_{m\mathbf{k}}}
f^{\lambda_2}_{n\mathbf{k+q},m\mathbf{k}}.
\end{split}
\end{equation}
It is useful to recall the following Hermiticity conditions for first-order occupation matrices 
(${\bf f}^\lambda$) 
and operators ($\hat{H}^\lambda$, $\hat{\mathcal{H}}^\lambda$ or $\hat{\rho}^\lambda$),
\begin{equation}
f^{\lambda \dagger}_{n\mathbf{k+q},m\mathbf{k}} = f^{\lambda}_{m\mathbf{k},n\mathbf{k+q}}, \qquad \hat{\rho}^{\lambda \dagger}_\mathbf{k,q} = \hat{\rho}^{\lambda}_\mathbf{k+q,-q},
\end{equation}
which guarantees that the Fourier transform of the response functions defined here are real.

The stationary conditions on $\bra*{u^{\lambda_1}_{m\mathbf{k,q}}}$ and
$f_{n\mathbf{k+q},m\mathbf{k}}^{\lambda_1\dagger}$ of this second-order energy functional give us the finite $\mathbf{q}$
counterparts of Eq. (\ref{Eq_f1}) and Eq. (\ref{Eq_Sternheimer}),
\begin{equation}\label{Eq_stationary_psi}
(\hat{H}^{(0)}_\mathbf{k+q}+a\hat{P}_\mathbf{k+q}-\epsilon_{m\mathbf{k}}^{(0)})
\ket*{u^{\lambda_2}_{m\mathbf{k,q}}}=-\hat{Q}_\mathbf{k+q}
\hat{\mathcal{H}}^{\lambda_2}_\mathbf{k,q}\ket*{u_{m\mathbf{k}}^{(0)}}
\end{equation}
and
\begin{equation}\label{Eq_stationary_f}
f^{\lambda_2}_{n\mathbf{k+q},m\mathbf{k}}=\frac{f_{n\mathbf{k+q}}-f_{m\mathbf{k}}}
{\epsilon^{(0)}_{n\mathbf{k+q}}-\epsilon^{(0)}_{m\mathbf{k}}}
\bra*{u_{n\mathbf{k+q}}^{(0)}}\hat{\mathcal{H}}^{\lambda_2}_\mathbf{k,q}
\ket*{u_{m\mathbf{k}}^{(0)}}.
\end{equation}
By plugging these two stationary conditions into the second-order energy functional, 
we obtain the following nonstationary expression,
\begin{equation}\label{Eq_E2_nonst_q}
E^{\lambda_1\lambda_2}_{\text{nonst},\mathbf{q}} = 
\int_\text{BZ}[d^3k] \, {\rm Tr}  \left( \hat{H}^{\lambda_1}_\mathbf{k+q,-q} \hat{\rho}^{\lambda_2}_\mathbf{k,q} \right),
\end{equation}
where the integrand is written in a compact form as a trace, and we have
introduced the first-order density operator,
\begin{equation}
\label{rholamq}
\begin{split}
 \hat{\rho}^{\lambda_2}_\mathbf{k,q} =&\sum_{m=1}^M 
 \Big( 
 \ket*{u_{m\mathbf{k,q}}^{\lambda_2}}f_{m\mathbf{k}}\bra*{u_{m\mathbf{k}}^{(0)}} \\
&+
\ket*{u_{m\mathbf{k+q}}^{(0)}} f_{m\mathbf{k+q}} \bra*{u_{m\mathbf{k+q,-q}}^{\lambda_2}} \Big)\\
&+\sum_{m,n=1}^M 
\ket*{u_{n\mathbf{k+q}}^{(0)}} f^{\lambda_2}_{n\mathbf{k+q},m\mathbf{k}}  \bra*{u_{m\mathbf{k}}^{(0)}}.
\end{split}
\end{equation} 

As outlined in Sec. \ref{Section_ensemble_E}, observable quantities must not
depend on the size of the active subspace. It can be readily demonstrated 
that the second-order energy functional, $E^{\lambda_1\lambda_2}_{\text{nonst},\mathbf{q}}$, is 
independent of $M$.
The proof relies on the $M$-independence of the first-order density operator, 
$\hat{\rho}^{\lambda_2}_\mathbf{k,q}$: if $M$ changes, part of the
spectral weight is transferred from the first two lines to the 
Kubo-like term in the third line of Eq.~(\ref{rholamq}), but the overall
sum remains unchanged.
In the limit where $M$ tends to infinity, the active space coincides with 
the entire Hilbert space; then, the first two lines vanish and the entire
operator is expressed in a Kubo-like sum-over-all-states (third 
line in Eq.~(\ref{rholamq})) form \cite{doi:10.1143/JPSJ.12.570}.
Conversely, in gapped systems at zero temperature, it is 
common practice to resctrict the active subspace 
to its bare minimum (i.e., to the valence manifold). In this case
$\Delta E^{\lambda_1\lambda_2}_\mathbf{k,q}$
vanishes identically, and the remainder contributions recover the 
well-known DFPT expressions for insulators~\cite{PhysRevB.55.10355,RevModPhys.73.515}.

\subsection{Time-reversal symmetry}\label{Sec_TR}

The formulas presented in the previous subsection have the drawback
that they require, in principle, to solve the Sternheimer problem 
simultaneously at ${\bf q}$ and ${\bf -q}$. 
In the following, we shall specialize our theory to crystals that
enjoy time-reversal (TR) symmetry, where this inconvenience can be 
avoided.
Indeed, assuming that both perturbations $\lambda_1$ and $\lambda_2$
are even under a TR operation, we have
\begin{equation}
\bra*{\mathbf{r}}\ket*{u_{m\mathbf{k}}^{(0)}} = \bra*{u_{m\mathbf{-k}}^{(0)}}\ket*{\mathbf{r}}, \quad
\bra*{\mathbf{r}}\ket*{u_{m\mathbf{k,q}}^{\lambda}} = 
\bra*{u_{m\mathbf{-k,-q}}^{\lambda}}\ket*{\mathbf{r}},
\end{equation}
which implies
\begin{equation}
\bar{E}^{\lambda_1\lambda_2 \, *}_\mathbf{k+q,-q} = \bar{E}^{\lambda_1\lambda_2}_\mathbf{-k-q,q}
\rightarrow \bar{E}^{\lambda_1\lambda_2}_\mathbf{k,q}.
\end{equation}
(Since the latter quantity must be anyway integrated over the Brillouin zone, 
we are allowed to operate an arbitrary shift in ${\bf k}$-space.)
This allows us to write the stationary expression for the second derivative
of the energy as
\begin{equation}\label{Eq_E2_TR}
\begin{split}
E^{\lambda_1\lambda_2}_{\text{st},\mathbf{q}}=&\int_\text{BZ}[d^3k]\left(
2\bar{E}_\mathbf{k,q}^{\lambda_1\lambda_2}
+\Delta E_\mathbf{k,q}^{\lambda_1\lambda_2}
\right)\\
&+\int_\Omega\int \rho^{\lambda_1 *}_\mathbf{q}(\mathbf{r})K_\mathbf{q}(\mathbf{r,r'})
\rho^{\lambda_2}_\mathbf{q}(\mathbf{r'}) 
d^3rd^3r' ,
\end{split}
\end{equation}
with the first-order electron densities defined as
\begin{equation}\label{Eq_rho1_TR}
\begin{split}
\rho_\mathbf{q}^\lambda(\mathbf{r})=\int_\text{BZ}[d^3k]\Bigg[
& 2 \sum_{m=1}^M f_{m\mathbf{k}}
\bra*{u_{m\mathbf{k}}^{(0)}}\ket*{\mathbf{r}}\bra*{\mathbf{r}}\hat{Q}_\mathbf{k+q}
\ket*{u_{m\mathbf{k,q}}^\lambda}\\
&+\sum_{m,n=1}^M
\bra*{u_{m\mathbf{k}}^{(0)}}\ket*{\mathbf{r}}\bra*{\mathbf{r}}\ket*{u_{n\mathbf{k+q}}^{(0)}}
f_{n\mathbf{k+q},m\mathbf{k}}^\lambda
\Bigg].
\end{split}
\end{equation}
Note that the first-order wave functions at ${\bf -q}$ are no longer needed.
After imposing the stationary principles, Eq. (\ref{Eq_stationary_psi}) and 
(\ref{Eq_stationary_f}), we arrive at an analogous 
nonstationary formula for the second derivative,
\begin{equation}\label{Eq_nonst_TR}
\begin{split}
E^{\lambda_1\lambda_2}_{\text{nonst},\mathbf{q}} = \int_\text{BZ}&[d^3k]\Bigg[
 2 \sum_{m=1}^M f_{m\mathbf{k}}
\bra*{u_{m\mathbf{k}}^{(0)}} ( \hat{H}^{\lambda_1}_\mathbf{k,q} )^\dagger
\ket*{u_{m\mathbf{k,q}}^{\lambda_2}}\\
&+\sum_{m,n=1}^M
\bra*{u_{m\mathbf{k}}^{(0)}} (\hat{H}^{\lambda_1}_\mathbf{k,q})^\dagger \ket*{u_{n\mathbf{k+q}}^{(0)}}
f_{n\mathbf{k+q},m\mathbf{k}}^{\lambda_2}
\Bigg].
\end{split}
\end{equation}

The use of TR symmetry in DFPT is, of course, well established. The reason why
we spell it out explicitly here is related to an important subtlety, specific 
to the metallic case, that is important to mention. 
The point is that, because of the shift in ${\bf k}$-space that we have operated 
on the TR-rectified ``${\bf -q}$'' terms, the \emph{integrands} (i.e., the quantities in
the square brackets) in Eq. (\ref{Eq_rho1_TR}) and (\ref{Eq_nonst_TR}) are no longer independent of
$M$: such property is restored only after integration over the full BZ is performed.
This observation has an undesirable consequence when operating the 
parametric differentiation in ${\bf q}$ (see next subsection): the accuracy
of the result depends on the vanishing of the total ${\bf k}$-derivative 
of $\bar{E}_\mathbf{k,0}^{\lambda_1\lambda_2}$. 
For this requirement to hold, we need $\bar{E}_\mathbf{k,0}^{\lambda_1\lambda_2}$
to be a differentiable function of ${\bf k}$, which is only true if the
active subspace forms an isolated group of bands (i.e., it must be separated 
from higher unoccupied states by a gap).

In all our test cases we were fortunate enough to find well-defined energy gaps in the conduction-band 
region of the spectrum, and we could always 
choose $M$ in such a way that
$\epsilon^{(0)}_{M\mathbf{k}}-\epsilon^{(0)}_{M+1\mathbf{k}}\neq0$ over the whole BZ.
Less fortunate cases might require choosing a different $M$ for distinct $\mathbf{k}$ points,
in order to avoid degeneracies between $M$ and $M+1$. (When such degeneracies are present, 
the linear-response Sternheimer problem becomes ill-conditioned, which makes it difficult or 
impossible to reach numerical convergence.)
Although in principle possible, doing so would require using Eq.~(\ref{Eq_E2_q}), in
place of Eq.~(\ref{Eq_E2_TR}) as a starting point for the long-wave expansion.
As this would entail a significant revision of the spatial-dispersion 
formulas already implemented in {\sc abinit}, we defer such developments to a future work.
\subsection{First-order in $\mathbf{q}$}\label{Sec_first_q}
In order to access spatial dispersion effects, our next task
consists in taking the analytical derivative of Eq. (\ref{Eq_E2_st_plus_nv})
with respect to ${\bf q}$.
Here the key advantage of our unconstrained variational formulation
becomes clear, as it allows for a straightforward application of the
$2n+1$ theorem.
At first order, in particular, we have \cite{PhysRevX.9.021050}
\begin{equation}
E^{\lambda_1\lambda_2}_\gamma=\frac{d E^{\lambda_1\lambda_2}_\mathbf{q}}
{d q_\gamma}\bigg|_\mathbf{q=0}=\frac{\partial E^{\lambda_1\lambda_2}_\mathbf{q}}
{\partial q_\gamma}\bigg|_\mathbf{q=0}.
\end{equation}
This means that the total derivative in $\mathbf{q}$ coincides 
with the partial derivative of the second-order energy functional, where the variables 
that are implicitly defined by the stationary principle (in our case \emph{both} the 
first-order wave functions, $\ket*{u^\lambda_{m\mathbf{k,q}}}$, 
and occupation matrices, $f^\lambda_{m\mathbf{k},n\mathbf{k+q}}$) are excluded from differentiation.
In other words, the time-consuming self-consistent solution of the Sternheimer problem 
only needs to be performed at ${\bf q}=0$, just like in the insulating case.
The only task we are left with consists in taking the derivatives of the nonvariational 
quantities in Eq.~(\ref{Eq_E2_q}) that explicitly depend on ${\bf q}$, i.e., the 
external potentials and the ground-state wave functions. 
The final result for the ${\bf q}$-derivative of the stationary part 
reads as
\begin{align}\label{Eq_E2_1q}
\begin{split}
E_{\text{st},\gamma}^{\lambda_1\lambda_2}=&\int_\text{BZ}[d^3k]\left(
2\bar{E}_{\mathbf{k},\gamma}^{\lambda_1\lambda_2}
+\Delta E_{\mathbf{k},\gamma}^{\lambda_1\lambda_2}
\right)\\
&+\int_\Omega \int \rho^{\lambda_1*}(\mathbf{r})
K_\gamma(\mathbf{r,r'})\rho^{\lambda_2}(\mathbf{r'})d^3rd^3r',
\end{split}
\end{align}
where 
\begin{equation}
K_\gamma(\mathbf{r,r'})=\frac{\partial K_\mathbf{q}(\mathbf{r,r'})}
{\partial q_\gamma}\Big|_\mathbf{q=0}
\end{equation}
represents the first $\mathbf{q}$-gradient of the Hxc kernel.
The wave-function term is in the exact same form as
in the insulating case \cite{PhysRevX.9.021050},
\begin{widetext}
\begin{equation}\label{Eq_wf_longwave}
\begin{split}
\bar{E}_{\mathbf{k},\gamma}^{\lambda_1\lambda_2}=&
\sum_{m=1}^M f_{m\mathbf{k}}\bra*{u_{m\mathbf{k}}^{\lambda_1}}
\hat{H}_\mathbf{k}^{k_\gamma}\ket*{u_{m\mathbf{k}}^{\lambda_2}}\\
&-\sum_{m,n=1}^{M}f_{m\mathbf{k}}\Big(
\bra*{u_{m\mathbf{k}}^{\lambda_1}}\ket*{u_{n\mathbf{k}}^{k_\gamma}}
\bra*{u_{n\mathbf{k}}^{(0)}}\hat{\mathcal{H}}_\mathbf{k}^{\lambda_2}
\ket*{u_{m\mathbf{k}}^{(0)}}+\bra*{u_{m\mathbf{k}}^{(0)}}
(\hat{\mathcal{H}}_\mathbf{k}^{\lambda_1})^\dagger
\ket*{u_{n\mathbf{k}}^{(0)}}\bra*{u_{n\mathbf{k}}^{k_\gamma}}
\ket*{u_{m\mathbf{k}}^{\lambda_2}}
\Big)\\
&+\sum_{m=1}^Mf_{m\mathbf{k}}\left(\bra*{u_{m\mathbf{k}}^{\lambda_1}}
\hat{H}_{\mathbf{k},\gamma}^{\lambda_2}
\ket*{u_{m\mathbf{k}}^{(0)}}+\bra*{u_{m\mathbf{k}}^{(0)}}
(\hat{H}_{\mathbf{k},\gamma}^{\lambda_1})^\dagger\ket*{u_{m\mathbf{k}}^{\lambda_2}}\right),
\end{split}
\end{equation}
and the occupation term, specific to metals, 
reads as
\begin{equation}\label{Eq_occ_longwave}
\begin{split}
\Delta E_{\mathbf{k},\gamma}^{\lambda_1\lambda_2}=&\sum_{m,n=1}^M\bar{f}_{mn\mathbf{k}}
\left(
\bra*{u_{m\mathbf{k}}^{(0)}}(\hat{\mathcal{H}}^{\lambda_1}_\mathbf{k})^\dagger
\ket*{u_{n\mathbf{k}}^{(0)}}\bra*{u_{n\mathbf{k}}^{(0)}}\hat{H}_{\mathbf{k},\gamma}^{\lambda_2}
\ket*{u_{m\mathbf{k}}^{(0)}}
+\bra*{u_{m\mathbf{k}}^{(0)}}(\hat{H}_{\mathbf{k},\gamma}^{\lambda_1})^\dagger
\ket*{u_{n\mathbf{k}}^{(0)}}\bra*{u_{n\mathbf{k}}^{(0)}}
\hat{\mathcal{H}}_\mathbf{k}^{\lambda_2}\ket*{u_{m\mathbf{k}}^{(0)}}
\right)\\
&+\sum_{m,n=1}^M\bar{f}_{mn\mathbf{k}}\left(
\bra*{u_{m\mathbf{k}}^{(0)}}(\hat{\mathcal{H}}_\mathbf{k}^{\lambda_1})^\dagger
\ket*{u_{n\mathbf{k}}^{(0)}}\bra*{u_{n\mathbf{k}}^{k_\gamma}}
\hat{\mathcal{H}}_\mathbf{k}^{\lambda_2}
\ket*{u_{m\mathbf{k}}^{(0)}}
+\bra*{u_{m\mathbf{k}}^{(0)}}(\hat{\mathcal{H}}_\mathbf{k}^{\lambda_1})^\dagger
\ket*{u_{n\mathbf{k}}^{k_\gamma}}\bra*{u_{n\mathbf{k}}^{(0)}}
\hat{\mathcal{H}}_\mathbf{k}^{\lambda_2}\ket*{u_{m\mathbf{k}}^{(0)}}
\right)\\
&+\sum_{m,n,l=1}^M\mathcal{F}_{mnl\mathbf{k}}
\bra*{u_{m\mathbf{k}}^{(0)}}(\hat{\mathcal{H}}_\mathbf{k}^{\lambda_1})^\dagger
\ket*{u_{n\mathbf{k}}^{(0)}}\bra*{u_{n\mathbf{k}}^{(0)}}
\hat{H}_\mathbf{k}^{k_\gamma}\ket*{u_{l\mathbf{k}}^{(0)}}
\bra*{u_{l\mathbf{k}}^{(0)}}\hat{\mathcal{H}}_\mathbf{k}^{\lambda_2}
\ket*{u_{m\mathbf{k}}^{(0)}}.
\end{split}
\end{equation}
\end{widetext}
The proof of Eq.~(\ref{Eq_occ_longwave}) rests on the rules
for the differentiation of operators outlined in Sec.~\ref{Sec_deriv_operators},
which we apply here to the case $\lambda = q_\gamma$ and $h(\e^{(0)}_{m{\bf k+q}}) =  
G(\e^{(0)}_{n\bf k},\e^{(0)}_{m{\bf k+q}})$.

\begin{figure}[b!]
	\includegraphics[width=1.0\linewidth]{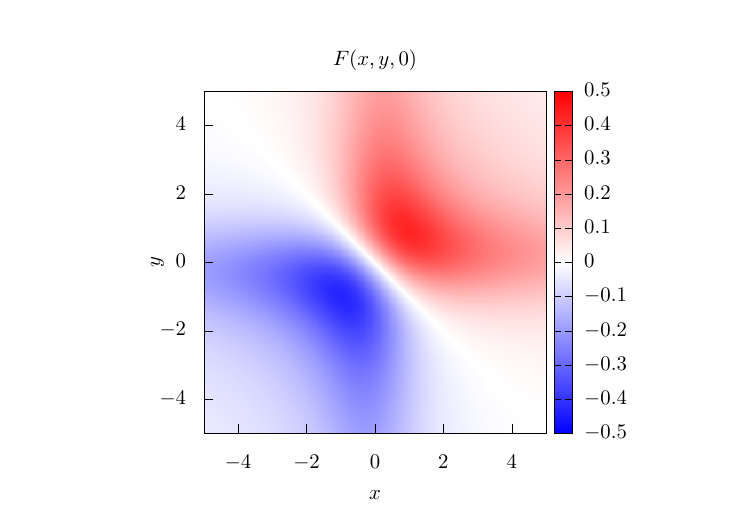}\\
	\includegraphics[width=1.0\linewidth]{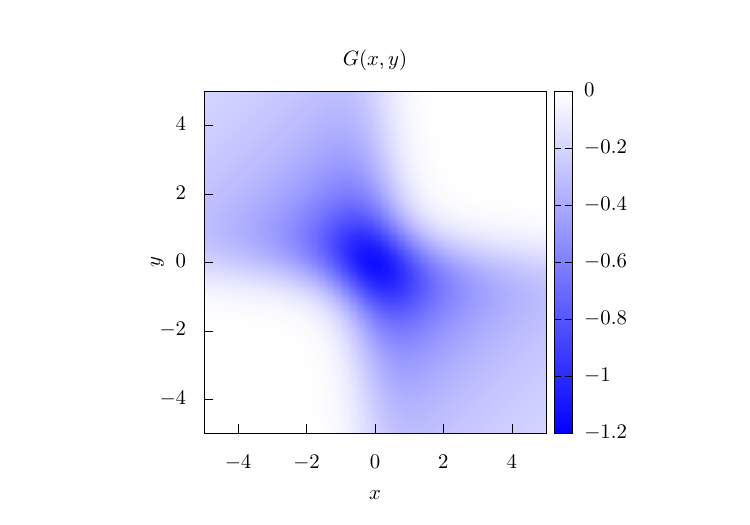}
	\caption{Two-dimensional contour plots (in arbitrary units) of the $\mathcal{F}$ and $G$ functions as defined in Eq. (\ref{Eq_calligraphic_F}) and Eq. (\ref{Eq_G}), respectively. The upper panel shows $\mathcal{F}(x,y,0)$ and the lower panel 
    $G(x,y)$.}
	\label{Fig_F}
\end{figure}

The new symbols appearing in Eq.~(\ref{Eq_wf_longwave}) and
Eq.~(\ref{Eq_occ_longwave}) are the $d/dk_\gamma$
wave functions, $\ket*{u_{m\mathbf{k}}^{k_\gamma}}$, the velocity operator,
\begin{equation}
\hat{H}_\mathbf{k}^{k_\gamma}=
\frac{\partial\hat{H}_\mathbf{k+q}^{(0)}}{\partial q_\gamma}\Big|_\mathbf{q=0}
\end{equation}
and
\begin{equation}
\hat{H}_{\mathbf{k},\gamma}^\lambda=
\frac{\partial\hat{H}^{\lambda}_\mathbf{k,q}}{\partial q_\gamma}\Big|_\mathbf{q=0}.
\end{equation}
The calligraphic symbol $\mathcal{F}_{mnl\mathbf{k}}\equiv
\mathcal{F}(\epsilon_{m\mathbf{k}},\epsilon_{n\mathbf{k}},\epsilon_{l\mathbf{k}})$, which is 
invariant under any permutation of the three band indices, is defined as \cite{PhysRevB.62.15283} 
\begin{equation}\label{Eq_calligraphic_F}
\mathcal{F}(x,y,z)=
\begin{cases}
\frac{G(x,y)-G(x,z)}{y-z}, \quad y\neq z\\[8pt]
\frac{1}{6}\left(\frac{\partial^2 f(x)}{\partial x^2}
+\frac{\partial^2 f(y)}{\partial y^2}
+\frac{\partial^2 f(z)}{\partial z^2}\right) ,\quad
x=y=z.
\end{cases}
\end{equation}
(See the parenthetical comment after Eq. (\ref{Eq_G}).)
We have carefully tested that our implementation of the function $\mathcal{F}(x,y,z)$ 
is continuous and smooth in the critical $x\simeq y \simeq z$ region. 
To illustrate the qualitative differences between $\mathcal{F}$ and $G$, 
in Fig. \ref{Fig_F} we present two filled contour plots  
of the functions $G(x,y)$ and $\mathcal{F}(x,y,0)$: their respective symmetric and antisymmetric 
nature under the interchange of the two arguments is clear.
In close analogy to the finite-${\bf q}$ case, $\Delta E_{\mathbf{k},\gamma}^{\lambda_1\lambda_2}$
vanishes in insulators if the active manifold is restricted to the occupied states; the present 
formalism reduces then to the already established results
\cite{PhysRevX.9.021050}.
\subsection{Treatment of the $\mathbf{q}\rightarrow \mathbf{0}$ limit (Fermi level shifts)}\label{Sec_Fermi_shifts}
Long-wave expansions generally require some care, as the Coulomb potential 
diverges in the ${\bf q}\rightarrow 0$ limit.
In insulators, this divergence results in a nonanalytic contribution to 
the response that is due to long-range electric fields. In metals, such fields are screened
by the redistribution of the free carriers, which tends to enforce local charge neutrality.
The requirement of charge neutrality becomes strict at ${\bf q}=0$, where 
it needs to be taken care of explicitly, via the holonomic constraint on
particle number in Eq. (\ref{Eq_ensemble_DFT_E}).

At second-order in the perturbations, this constraint  
propagates to the second-order energy functional as
$\mu^\lambda\Tr(\mathbf{f^\lambda})$, where $\mu^\lambda$ plays the role of a first-order Lagrange 
multiplier \cite{PhysRevA.52.1096}.  
By imposing the stationary condition on the $\mathbf{f^\lambda}$ matrix, one readily obtains
the so-called ``Fermi level shifts'' contribution,
\begin{equation}\label{Eq_f1_Fermi_shift}
\begin{split}
f^\lambda_{nm\mathbf{k}}&=\bar{f}_{nm\mathbf{k}}
\left(
\bra*{u^{(0)}_{n\mathbf{k}}}\hat{\mathcal{H}}^\lambda_\mathbf{k}
\ket*{u^{(0)}_{m\mathbf{k}}}-\delta_{mn}\mu^\lambda\right)\\
&=\bar{f}_{nm\mathbf{k}}
\bra*{u^{(0)}_{n\mathbf{k}}}
\big(
\hat{\mathcal{H}}^\lambda_\mathbf{k}-\mu^\lambda
\big)
\ket*{u^{(0)}_{m\mathbf{k}}}.
\end{split}
\end{equation}
A closed expression for $\mu^\lambda$ is can be obtained by imposing that the trace of 
$\mathbf{f}^\lambda$ should vanish,
\begin{equation}
\mu^\lambda=\frac{\displaystyle\int_\text{BZ}[d^3k]\sum_m f'_{m\mathbf{k}}
\bra*{u^{(0)}_{m\mathbf{k}}}\hat{\mathcal{H}}^\lambda_\mathbf{k}
\ket*{u^{(0)}_{m\mathbf{k}}}}
{\displaystyle\int_\text{BZ}[d^3k]\sum_m f'_{m\mathbf{k}}},
\end{equation}
where $f'_{m\mathbf{k}}=\frac{\partial f(\epsilon)}{\partial\epsilon}
\big|_{\epsilon=\epsilon^{(0)}_{m\mathbf{k}}}$.

It is important to stress that the Fermi-shift-corrected second-order 
energy at $\Gamma$ is the exact ${\bf q}\rightarrow 0$ limit of the second-order 
energy calculated at a small but finite ${\bf q}$; in the latter, the diverging 
${\bf G}=0$ term in the Coulomb kernel is still present, and therefore, no correction 
is needed. 
Note that the limit is the same regardless of the direction, which is
equivalent to observing that adiabatic response functions in metals 
at finite electronic temperature are always analytic functions of
${\bf q}$. 
This means that, unlike in insulators, the long-wave expansion of 
the force-constant matrix can be carried out on the whole response
function, without the need of artificially suppressing the 
macroscopic electric field contribution prior to differentiation.

This also implies that special care is needed at correctly differentiating 
such macroscopic electrostatic term with respect to ${\bf q}$ in 
Eq. (\ref{Eq_E2_st_plus_nv}).
The macroscopic (mac) electrostatic contributions are originated from three 
different sources: the first contribution comes from the nonvariational term,
the second contribution originates from the second line of 
Eq. (\ref{Eq_E2_q}), where the Coulomb kernel appears with the first-order electron densities, and  
the third term arises from the first line of the occupation term, Eq. (\ref{Eq_occ_longwave}).
All the divergences nullify each other, yielding 
an additional contribution that is proportional to the Fermi level shifts ($\mu^\lambda$) produced by the perturbations. For the phonon case,
\begin{equation}\label{Eq_correction_Fermi_shift_deriv}
E_{\text{mac},\gamma}^{\tau_{\kappa\alpha}\tau_{\kappa'\beta}}\simeq 
i\delta_{\alpha\gamma}Z_\kappa \mu^{\tau_{\kappa'\beta}}
-i\delta_{\beta\gamma}\mu^{\tau_{\kappa\alpha}}Z_{\kappa'},
\end{equation}
where $Z_\kappa$ is the bare nuclear pseudo-charge of sublattice $\kappa$.
(More details can be found in Appendix \ref{Appendix_electrostatics}.)

Since there is no need to remove the (ambiguous) macroscopic fields contributions
prior to the long-wave expansion, the adiabatic spatial dispersion coefficients are well-defined 
bulk properties in metals.
In other words, the far-away surfaces cannot contribute electrostatically to the bulk response, and
the problematic \textit{potential energy reference}~\cite{PhysRevB.105.064101} issue inherent 
to spatial dispersion in insulators is absent in metals.
\section{Application to phonons and strain gradients}\label{Sec_flexocoupling}
Up to now, our theory has been presented in a completely 
general form, and is valid as it stands for any pair of perturbations 
$\lambda_1$ and $\lambda_2$. 
For a numerical validation of the formal results presented thus far, 
in the following we shall focus our attention to phonon and strain gradients. 
These applications, while primarily intended
as a numerical illustration of our methodology,
have an obvious practical relevance as well, given
the ongoing interest in ferroelectric metals.

The basic ingredient for what follows is the force-constant (FC) matrix,
\begin{equation}\label{Eq_FC_q}
\begin{split}
\Phi^\mathbf{q}_{\kappa\alpha,\kappa'\beta}=&\sum_l \Phi^l_{\kappa\alpha,\kappa'\beta}
e^{i\mathbf{q}\cdot(\mathbf{R}^l+\boldsymbol{\tau}_{\kappa'}-\boldsymbol{\tau}_\kappa)},\\
\Phi^l_{\kappa\alpha,\kappa'\beta}=&\frac{\partial^2 E}
{\partial u^0_{\kappa\alpha}\partial u^l_{\kappa'\beta}},
\end{split}
\end{equation}
defined as the second derivative of the total energy $E$ with respect to
monochromatic lattice displacements of the type
\begin{equation}
u^l_{\kappa\alpha} = u_\alpha e^{i\mathbf{q}\cdot\mathbf{R}^l_{\kappa}},
\end{equation}
where $u_\alpha$ represents the deformation amplitude and
$\mathbf{R}^l_{\kappa}=\mathbf{R}^l+\boldsymbol{\tau}_\kappa$,
where $\mathbf{R}^l$ is a Bravais lattice vector of the $l$-th cell, and 
$\boldsymbol{\tau}_\kappa$ represents
the equilibrium position of the sublattice $\kappa$ within the unit cell; $\alpha$ and $\beta$ are Cartesian directions.
We write its long-wave expansion 
in a vicinity of ${\bf q}=0$ as \cite{Born1954,PhysRevB.88.174106}
\begin{equation}\label{Eq_FC}
\Phi^\mathbf{q}_{\kappa\alpha,\kappa'\beta}\simeq \Phi^{(0)}_{\kappa\alpha,\kappa'\beta}
-iq_\gamma \Phi^{(1,\gamma)}_{\kappa\alpha,\kappa'\beta}
-\frac{q_\gamma q_\delta}{2}\Phi^{(2,\gamma\delta)}_{\kappa\alpha,\kappa'\beta},
\end{equation}
were $\boldsymbol{\Phi}^{(0)}$ is the zone-center FC matrix, and $\boldsymbol{\Phi}^{(1)}$
and $\boldsymbol{\Phi}^{(2)}$ describe its spatial dispersion at first and second 
order in the momentum ${\bf q}$.
We shall discuss their physical relevance (and their relation to the theory 
developed insofar) separately in the following.

\subsection{Phonons}
$\Phi^{(1,\gamma)}_{\kappa\alpha,\kappa'\beta}$ has to do with the force produced on sublattice $\kappa$ along $\alpha$ by
a displacement pattern of the sublattice $\kappa'$ that is linearly increasing in space along $r_\gamma$.
If the lattice Hamiltonian were local (i.e., if the atomic lattice behaved like an array of noninteracting 
harmonic oscillators), such force would trivially correspond to $R_{\kappa'\gamma} \Phi^{(0)}$; $\boldsymbol{\Phi}^{(1)}$
describes the correction to that value that is due to the nonlocality of the interatomic forces.
Historically, $\boldsymbol{\Phi}^{(1)}$ was first introduced in the context of bulk flexoelectricity, where it
mediates an indirect contribution to the lattice response to a macroscopic strain gradient~\cite{PhysRevB.88.174106,PhysRevB.105.064101}.
Such a contribution is relevant whenever the crystal allows for Raman-active lattice modes,
e.g., in diamond-structure crystals (bulk Si or C) and tilted perovskites like SrTiO$_3$ or LiOsO$_3$.
More recently, its importance was pointed out in ferroic crystals, where it acquires a central place in the context of
nonlinear gradient couplings (most notably, in the form of antisymmetric Dzyaloshinskii-Moriya-like terms \cite{stengel2023macroscopic})
between lattice modes.
In chiral crystals such as $\alpha$-HgS,  
the main physical consequence of $\boldsymbol{\Phi}^{(1)}$
consists in the appearence of chiral phonon modes with opposite angular 
momentum that disperse linearly along the main crystal axis 
\cite{ishito2023truly}.
Such an effect can be regarded as the phonon counterpart of 
the natural optical activity \cite{landau1984electrodynamics}.

Based on the theory developed thus far, in combination with the
established methodology that was already developed for the insulating case~\cite{PhysRevB.105.064101},
we calculate $\boldsymbol{\Phi}^{(1)}$ as
\begin{equation}
\Phi^{(1,\gamma)}_{\kappa\alpha,\kappa'\beta}=-\Im\left(
E_{\text{st},\gamma}^{\tau_{\kappa\alpha}\tau_{\kappa'\beta}}
+E_{\text{Ew},\gamma}^{\tau_{\kappa\alpha}\tau_{\kappa'\beta}}
\right).
\end{equation}
The stationary part is straightforwardly obtained by substituting 
$\lambda_1=\tau_{\kappa\alpha}$ and $\lambda_2=\tau_{\kappa'\beta}$ into 
Eq. (\ref{Eq_E2_1q}),
while the nonvariational contribution acquires the form of the first $\mathbf{q}$ derivative of the ionic
Ewald (Ew) energy, whose explicit expression
can be found in Appendix A of Ref. \cite{PhysRevB.105.064101}.

\subsection{Strain gradients}\label{Sec_strain_gradients}
$\boldsymbol{\Phi}^{(2)}$ (third term in the right-hand side of Eq. (\ref{Eq_FC})), on the other hand, 
is directly related to the forces on individual sublattices produced by a strain gradient, i.e., the so-called \emph{flexoelectric force-response} tensor.
Its type-I representation, indicated by the \textit{square bracket}
symbol, can be written as
\begin{equation}\label{Eq_square}
[\alpha\beta,\gamma\delta]^\kappa =-\frac{1}{2}\sum_{\kappa'}
\Phi^{(2,\gamma\delta)}_{\kappa\alpha,\kappa'\beta}.
\end{equation}
(A type-II strain gradient refers to the gradient of a sym-
metrized strain, while its type-I counterpart describes the
response to an unsymmetrized strain gradient.)
Switching from type-I to type-II representation (and vice versa) is straightforward, through
the subsequent rearrangement of the indices \cite{Born1954,PhysRevB.88.174106},
\begin{equation}\label{Eq_C_square}
\bar{C}^\kappa_{\alpha\gamma,\beta\delta}=
[\alpha\beta,\gamma\delta]^\kappa
+[\alpha\delta,\beta\gamma]^\kappa
-[\alpha\gamma,\beta\delta]^\kappa,
\end{equation}
where the ``bar" symbol indicates that we are studying the response at the clamped-ion level.
One of the most notable physical manifestation of the latter is the flexocoupling tensor,
which accounts for the forces on the zone-center polar modes of the system produced 
by the applied strain gradient, and can be expressed in the following way,
\begin{equation}\label{Eq_flexocoupling}
f_{\alpha\lambda,\beta\gamma}=\sum_{\kappa,\alpha}\sqrt{\frac{M}{m_\kappa}}P^{(\alpha)}_{\kappa\rho}
C^\kappa_{\rho\lambda,\beta\gamma},
\end{equation}
where $m_\kappa$ is the mass of the sublattice $\kappa$, $M=\sum_\kappa m_\kappa$ is the total mass
of the unit cell and $\mathbf{P}^{(\alpha)}$ is a normalized polar eigenvector of the 
zone-center dynamical matrix. (The factor $\sqrt{M/m_\kappa}$ in 
Eq. (\ref{Eq_flexocoupling}) follows the convention of earlier works \cite{PhysRevB.93.245107,PhysRevX.12.031002,PhysRevLett.126.127601}.)

Directly applying the formulas of Sec. \ref{Sec_first_q}
to the calculation of $\boldsymbol{\Phi}^{(2)}$ is not possible, as they exclusively target first order terms in ${\bf q}$.
However, note that Eq. (\ref{Eq_square}) only requires the sublattice sum of the 
$\boldsymbol{\Phi}^{(2)}$ coefficients,
which physically corresponds to the force-response to an \emph{acoustic phonon} perturbation. 
Acoustic phonons can be conveniently recast, via a coordinate transformation to the comoving frame~\cite{PhysRevB.98.125133},
to a metric-wave perturbation~\cite{PhysRevB.99.085107}. This allows us to write the flexoelectric force-response
coefficients as first-order dispersion of the piezoelectric force-response tensor~\cite{PhysRevB.105.064101}.
In light of this, the whole story boils down to applying the theory developed 
in Sec. \ref{Sec_first_q} to the case in which $\lambda_1=\tau_{\kappa\alpha}$ and
$\lambda_2=\eta_{\beta\delta}$, 
where $\boldsymbol{\eta}$ denotes the uniform strain perturbation~\cite{PhysRevB.71.035117}.
In practice, the type-II representation of the flexoelectric force-response tensor
exhibits the following formulation~\cite{PhysRevB.105.064101},
\begin{equation}
\bar{C}^\kappa_{\alpha\gamma,\beta\delta}=
E^{\tau_{\kappa\alpha}\eta_{\beta\delta}}_{\text{st},\gamma}
+E^{\tau_{\kappa\alpha}\eta_{\beta\delta}}_{\text{nv},\gamma}.
\end{equation}
(Our notation slightly differs from Refs. \cite{PhysRevX.9.021050,PhysRevB.105.064101}, where $(\beta\delta)$ is used instead of $\eta_{\beta\delta}$.)
Remarkably, the nonvariational contribution takes the exact same form found in insulators
\cite{PhysRevB.105.064101}, the only difference being that, instead of assuming that all the 
active states are completely filled (this is the case in insulators as the active subspace is usually restricted to the valence manifold), the occupation function 
$f_{m\mathbf{k}}$ must be taken into account for each band $m$ and $\mathbf{k}$ point.
The stationary part is given by
\begin{equation}\label{Eq_C_st}
\begin{split}
E_{\text{st},\gamma}^{\tau_{\kappa\alpha}(\beta\delta)}=&
\int_\text{BZ}[d^3k]
\left(
2\bar{E}_{\mathbf{k},\gamma}^{\tau_{\kappa\alpha}\eta_{\beta\delta}}
+\Delta E_{\mathbf{k},\gamma}^{\tau_{\kappa\alpha}\eta_{\beta\delta}}
\right)\\
&+i\int_\Omega\int \rho^{\tau_{\kappa\alpha}*}(\mathbf{r})K_\gamma(\mathbf{r,r'})
\rho^{\eta_{\beta\delta}}(\mathbf{r'})d^3rd^3r',
\end{split}
\end{equation}
where $\eta_{\beta\delta}$ represents a uniform strain perturbation, as formulated by
Hamann \textit{et at.}
(HWRV) \cite{PhysRevB.71.035117}.
\subsection{Sum rules}
In order to validate our computational strategy for the calculation of the 
$\boldsymbol{\Phi}^{(1)}$ and $\boldsymbol{\bar{C}}^\kappa$ tensors,
it is useful to recall the following
well-established relationships \cite{Born1954},
\begin{equation}\label{Eq_sum_rule_Phi}
\Lambda^\kappa_{\alpha\beta\gamma}=\sum_\kappa \Phi^{(1,\gamma)}_{\kappa\alpha,\kappa'\beta}+f_{\kappa\beta}\delta_{\alpha\gamma}
\end{equation}
and
\begin{equation}\label{Eq_sum_rule_C}
\begin{split}
\bar{C}^\text{HWRV}_{\alpha\gamma,\beta\delta}=\frac{1}{\Omega}\sum_\kappa \bar{C}^\kappa_{\alpha\gamma,\beta\delta}+
\frac{1}{2}\Big(
\delta_{\gamma\beta}S_{\alpha\delta}+\delta_{\gamma\delta}S_{\alpha\beta}\\
-\delta_{\alpha\beta}S_{\gamma\delta}-\delta_{\alpha\delta}S_{\gamma\beta}
+2\delta_{\alpha\gamma}S_{\beta\delta}
\Big),
\end{split}
\end{equation}
where $\Lambda^\kappa_{\alpha\beta\gamma}$ and $\bar{C}^\text{HWRV}_{\alpha\gamma,\beta\delta}$ are,
respectively, 
the piezoelectric force-response tensor and the clamped-ion macroscopic elastic tensor, which 
are routinely computed in standard DFT codes with the metric-tensor formulation 
as proposed by HWRV \cite{PhysRevB.71.035117}. $f_{\kappa\beta}$
represents the atomic force on the sublattice $\kappa$ along the Cartesian direction $\beta$ and 
$S_{\alpha\delta}$ is the stress tensor, which is symmetric under the exchange 
$\alpha\leftrightarrow\delta$. While the sum rules  provided by Eq. (\ref{Eq_sum_rule_Phi}) and 
(\ref{Eq_sum_rule_C}) were initially established for insulators \cite{PhysRevB.105.064101},
in Sec. \ref{Sec_results} we numerically prove their validity in metal materials, demonstrating that 
no additional modifications are required. (Unless otherwise stated, we shall assume that the system under study is at mechanical equilibrium, i.e., forces and stresses tend to zero.)
\section{Results}\label{Sec_results}
\subsection{Computational parameters}
\begin{figure}[t!]
	\includegraphics[width=1.0\linewidth]{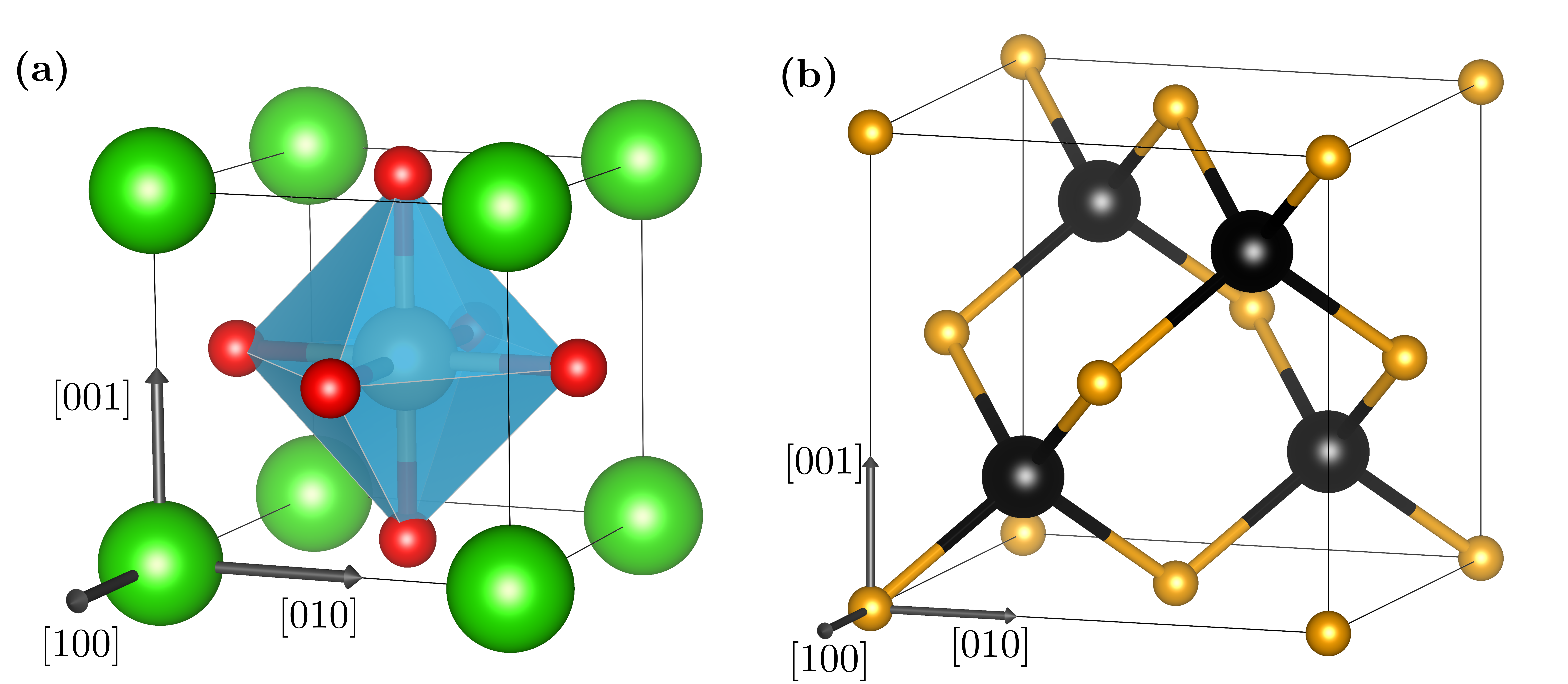}
	\caption{Structures of (a) cubic LiOsO$_3$ and (b) TiB. Grey arrows indicate
		the crystallographic directions that form the Cartesian axes. }
	\label{Fig_LiOsO3}
\end{figure}

We shall consider two different crystal structures.
First, we shall study the zincblende-type structure metals TiB and SiP, which constitute valuable theoretical models for testing our implementation. 
The selection of the face-centered cubic (fcc) structures for TiB and SiP, which belong to the space group $F\bar{4}3m$ and contain two atoms per unit cell, is motivated by their high symmetry and their structural resemblance to zincblende, an extensively studied compound.
A schematic representation of the TiB crystal is given in 
Fig. \ref{Fig_LiOsO3} (b). Second, we shall draw our attention to
LiOsO$_3$, a polar metal that has been attracting a lot of interest over the last few years, 
particularly after its experimental observation in 2013 \cite{shi2013ferroelectric}. For simplicity,
we shall restrict our analysis to the cubic phase.
A cartoon representing its unit cell is depicted in 
Fig. \ref{Fig_LiOsO3} (a).

Our first principles calculations are performed with the 
DFT and DFPT implementations of the open-source {\sc abinit} \cite{GONZE2020107042,GONZE20092582} package with the Perdew-Wang \cite{PhysRevB.45.13244} parametrization of the local density
approximation (LDA). To facilitate a comparison with earlier works, we also 
employed the 
Perdew-Burke-Ernzerhof (PBE) \cite{PhysRevLett.77.3865} parametrization of the generalized
gradient approximation (GGA) for lithium osmate.
Our long-wave ensemble DFPT expressions for the computation of dispersion properties in metals, Eq. (\ref{Eq_E2_1q}) to (\ref{Eq_occ_longwave}), are incorporated to the
{\sc abinit} package after minor modifications to the 
recently implemented longwave module. Norm-conserving pseudopotentials from the
Pseudo Dojo \cite{VANSETTEN201839} website are used as input to the 
ONCVPSP \cite{PhysRevB.88.085117} software, in order to regenerate them
without exchange-correlation nonlinear core corrections.

All our first-principles calculations are carried out with a plane-wave cutoff of 60 Ha, 
a Gaussian smearing of $0.01$ Ha
and 
the BZ is sampled with a dense Monkhorst-Pack mesh of $20\times20\times20$ $\mathbf{k}$ points. The crystal structures are relaxed until all the forces are smaller than $0.5\times10^{-7}$ Ha/bohr, obtaining a unit cell parameter of $a=9.176$ bohr for TiB and $a=9.879$ bohr for SiP. For LiOsO$_3$, we obtain a
relaxed cell parameter of $a=7.132$ bohr with LDA and $a=7.253$ bohr with GGA.

The active subspace is chosen to be either $M=10$ or $M=14$ for TiB, $M=8$ for SiP and $M=20$ for 
LiOsO$_3$. 
These choices guarantee that the active subspace forms an isolated group of bands in all
cases, following the observations in the last paragraph of Sec. \ref{Sec_TR}.
(Using two different values for $M$ in the case of TiB allows us to test the 
consistency of our implementation, and more specifically the independence 
of the converged results on the dimension of the active subspace.)
\subsection{TiB and SiP}
\begin{figure}[b!]
	\includegraphics[width=1.0\linewidth]{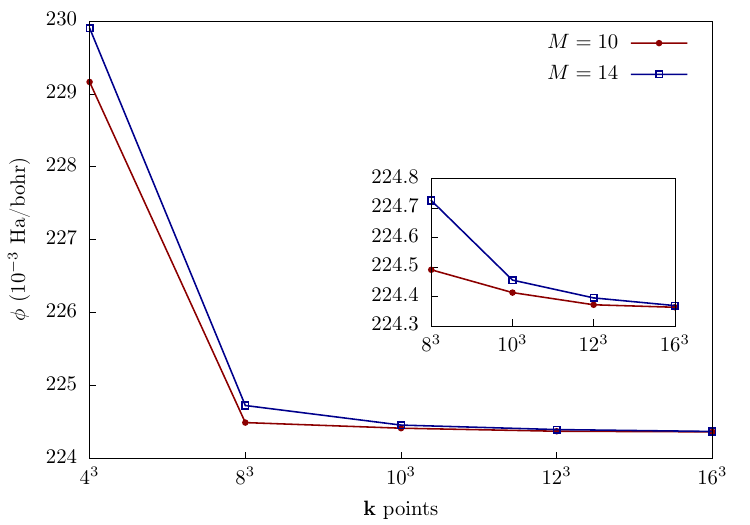}
	\caption{Convergence of the $\phi$ parameter (see Eq. (\ref{Eq_Phi_Lambda_indep})) for TiB as a function of the
	$\mathbf{k}$ point mesh resolution, for $M=10$ and $M=14$. This is a numerical validation that, as long as the highest states considered in the calculation have vanishing occupations, all the observables that can be extracted from the energy functional are independent of the size of the active subspace, $M$. Solid lines are a guide to the eye.}
	\label{Fig_iB}
\end{figure}

We shall start by 
testing our implementation with the 
two simple metals TiB and SiP. 
The symmetries of the materials substantially reduce the number of independent components of $\boldsymbol{\Phi}^{(1)}$ and $\boldsymbol{\bar{C}}^\kappa$.
For example, Eq. (\ref{Eq_sum_rule_Phi}) reduces 
to $\phi=\lambda$, since \cite{PhysRevB.105.064101}
\begin{equation}\label{Eq_Phi_Lambda_indep}
\begin{split}
\Phi^{(1,\gamma)}_{\kappa\alpha,\kappa'\beta}&=(-1)^{\kappa+1}(1-\delta_{\kappa\kappa'})
\phi\abs{\epsilon_{\alpha\beta\gamma}},\\
\Lambda^\kappa_{\alpha\beta\gamma}&=(-1)^{\kappa+1}\lambda\abs{\epsilon_{\alpha\beta\gamma}},
\end{split}
\end{equation}
where $\epsilon_{\alpha\beta\gamma}$ is the Levi-Civita symbol. 
In Table \ref{TiB_and_SiP_lw_2} we show the independent components of the (type-II)
flexoelectric force-response tensor, and Table \ref{TiB_and_SiP_lw_1} shows 
the only independent component of $\boldsymbol{\Phi}^{(1)}$ 
($\boldsymbol{\Lambda}$), indicated 
as $\phi$ ($\lambda$). The sum rules Eq. (\ref{Eq_sum_rule_Phi}) and Eq. (\ref{Eq_sum_rule_C})
are validated to a remarkably high level of accuracy. 
\begin{table}[t!]
	\begin{ruledtabular}
		\caption{Linearly independent components of the flexoelectric 
			force-response tensor (in eV) and the clamped-ion elastic tensor (in GPa) for 
			TiB and SiP.
			The latter is computed via the sublattice sum of 
			$\bar{C}^\kappa_{\alpha\gamma,\beta\delta}$ as indicated by
			Eq. (\ref{Eq_sum_rule_C}), and with the standard HWRV implementation.}
		\begin{tabular}{cc|rrr}
			&&$xx,xx$& $xx,yy$ & $xy,xy$\\ \hline
			\multirow{4}*{TiB}& $\bar{C}^\text{Ti}_{\alpha\gamma,\beta\delta}$&
			5.503&10.373&7.110\\
			&$\bar{C}^\text{B}_{\alpha\gamma,\beta\delta}$&
			10.225&12.163&6.428\\
			& $\bar{C}_{\alpha\gamma,\beta\delta}$ &88.108&126.250 &75.843\\
			& $\bar{C}^\text{HWRV}_{\alpha\gamma,\beta\delta}$&88.069 &126.186 &75.812\\ \hline
			\multirow{4}*{SiP}& $\bar{C}^\text{Si}_{\alpha\gamma,\beta\delta}$&
			8.542&14.861 &7.034\\
			& $\bar{C}^\text{P}_{\alpha\gamma,\beta\delta}$&
			1.572&7.911&3.676\\
			& $\bar{C}_{\alpha\gamma,\beta\delta}$ &45.373 &102.162 &48.049\\
			&$\bar{C}^\text{HWRV}_{\alpha\gamma,\beta\delta}$&45.615&102.223&48.260	      
		\end{tabular}
		\label{TiB_and_SiP_lw_2}
	\end{ruledtabular}
\end{table}
\begin{table}[t!]
	\begin{ruledtabular}
		\caption{A comparison between $\phi$ and $\lambda$, in $10^{-3}$ Ha/bohr units 
			(see Eq. \ref{Eq_Phi_Lambda_indep}). $\phi$ is computed with our long-wave ensemble
		DFPT formalism presented in this work and $\lambda$ with the standard HWRV implementation.}
		\begin{tabular}{c|cc}
	          &$\phi$& $\lambda$  \\ \hline
		 TiB  &224.360 &224.368   \\
		 SiP  &287.126 &287.340  	 	      		 	      
		\end{tabular}
		\label{TiB_and_SiP_lw_1}
	\end{ruledtabular}
\end{table} 
As an additional test of our implementation, we show in Fig. \ref{Fig_iB}
the computed $\phi$ parameter for TiB as a function of the
$\mathbf{k}$ points mesh resolution. Furthermore, in order to prove that results remain unaltered
irrespective of variations in the parameter $M$ (size of the active subspace), we show a comparison between $M=10$ and $M=14$. The obtained results reveal a high level of agreement, which corroborates the robustness of our implementation.
(The discrepancies between the results obtained with $M=10$ and $M=14$ in Fig. \ref{Fig_iB} for small $\mathbf{k}$ points samplings can be attributed to the shift in $\mathbf{k}$ space that we discussed in Sec. \ref{Sec_TR}. Had we adhered to Eq. (\ref{Eq_E2_q}), instead of utilizing Eq. (\ref{Eq_E2_TR}), the results in Fig. \ref{Fig_iB} would have aligned perfectly, regardless of the number of $\mathbf{k}$ points employed in the calculation.)

To conclude with the zincblende structures and in order to validate the sum rules 
Eq. (\ref{Eq_sum_rule_Phi}) and (\ref{Eq_sum_rule_C}) in presence of nonvanshing forces and stresses, we apply a displacement of 0.3 bohr along the $x$ Cartesian direction to the B atom in TiB.
The resulting crystal structure belongs to the space group $Imm2$ and we obtain, in absolute value, maximum 
interatomic forces of $3\times10^{-2}$ Ha$/$bohr and stress components of
$3.5\times 10^{-4}$ Ha$^3/$bohr.
Table \ref{TiB_forces} and \ref{TiB_stress} show selected tensor elements
of $\boldsymbol{\Phi}^{(1)}$ and 
$\boldsymbol{\bar{C}}$, respectively.
\begin{table}[t!]
	\begin{ruledtabular}
		\caption{Independent components of the piezoelectric force-response tensor of distorted TiB, computed
		with the standard HWRV implementation and with the sum rule of Eq. (\ref{Eq_sum_rule_Phi}).
		The prime symbol indicates that the contribution coming from the interatomic forces has not been taken into account. Values are given in $10^{-3}$ Ha/bohr.}
		\begin{tabular}{crrr}
			&HWRV& [Eq. (\ref{Eq_sum_rule_Phi})]$'$  & Eq. (\ref{Eq_sum_rule_Phi})\\ \hline
			$\Lambda^\text{Ti}_{xxx}$  & 64.420& 35.140 & 64.417\\
			$\Lambda^\text{Ti}_{xyy}$  & $-$104.004 & $-$104.876 &$-$104.876 \\
			$\Lambda^\text{Ti}_{xzy}$  &   $-$221.380 & $-$220.530 & $-$220.530 \\
			$\Lambda^\text{Ti}_{yxy}$  & $-$49.769&  $-$78.171&$-$48.894\\   
			$\Lambda^\text{Ti}_{yzx}$  & $-$231.073& $-$231.070&$-$231.070\\    
		\end{tabular}
		\label{TiB_forces}
	\end{ruledtabular}
\end{table} 
\begin{table}[t!]
	\begin{ruledtabular}
		\caption{Selected clamped-ion elastic tensor coefficients of distorted TiB, computed
			with the standard HWRV implementation and with the sum rule of Eq. (\ref{Eq_sum_rule_C}).
			The prime symbol indicates that the contributions coming from the stress have not been taken into account. Values are given in GPa units.}
		\begin{tabular}{crrr}
			&$\bar{C}^\text{HWRV}_{\alpha\gamma,\beta\delta}$& [Eq. (\ref{Eq_sum_rule_C})]$'$  & Eq. (\ref{Eq_sum_rule_C})\\ \hline
			$xx,xx$  & 78.258 & 76.822  & 78.293\\ 
			$xx,yy$  & 122.877& 125.366  & 122.943\\   
			$xx,yz$  & 19.023 & 29.251  &19.038 \\    
			$xy,xy$  & 71.972& 70.060  & 72.007\\
			$xy,xz$  & 17.575&  12.478&17.584 \\
			$yx,xy$  & 71.972& 73.958 & 72.011\\
			$yx,xz$  & 17.575& 22.695 & 17.589\\
			$yy,xx$  & 122.877& 121.470 & 122.941\\
			$yy,yy$  & 87.762&90.233& 87.810\\
			$yy,yz$  & 31.655 &41.900& 31.687\\
			$yy,zz$  & 135.424&137.935& 135.512\\
			$yz,yy$  & 31.655&21.449& 31.662\\
		\end{tabular}
		\label{TiB_stress}
	\end{ruledtabular}
\end{table} 
\subsection{Cubic LiOsO$_3$}
In the cubic phase of LiOsO$_3$ all the atoms sit at inversion centers, resulting in a vanishing 
$\boldsymbol{\Phi}^{(1)}$ tensor. This system represents, however, the ideal scenario for testing 
our implementation in the context of strain gradients, as the flexocoupling coefficients of cubic lithium osmate have been 
previously computed in Ref. \cite{PhysRevLett.126.127601}.
Table \ref{table:cubic_20k} presents the calculated independent 
components for the flexoelectric force-response tensor, along with the numerical validation of Eq. (\ref{Eq_sum_rule_C}). 
Additionally, the last two rows of Table \ref{table:cubic_20k} display the independent components of the flexocoupling tensor, computed by means of Eq. (\ref{Eq_flexocoupling}). The penultimate row shows the numerical values obtained in the present work, whereas the last row 
corresponds to the results from Ref. \cite{PhysRevLett.126.127601}.

The agreement between the present results with those of 
Ref.  \cite{PhysRevLett.126.127601} is remarkable, especially considering the different computational strategy 
employed therein. (In Ref. \cite{PhysRevLett.126.127601}, the flexocoupling coefficients were calculated 
via lattice sums of the real-space interatomic FCs, as opposed to the analytical long-wave approach
presented here.)
Regarding the effect of different schemes for the exchange-correlation functional,
we observe that the equilibrium volume slightly differs between LDA and GGA, following the expected trends: we obtain 
$\Omega=362.756$ bohr$^3$ and $\Omega=381.532$ bohr$^3$, respectively. 
In general, the calculated values for the flexoelectric force-response tensor and 
the flexocoupling coefficients show only a small dependence on the choice of
the exchange and correlation model.
It is also worth mentioning that the limitations of the long-wave module as implemented in the v9 version of {\sc abinit} 
prohibits the use of pseudopotentials that include non-linear core corrections. The small
disparities for the elastic constants and the flexocoupling tensor obtained with the PBE 
parametrization of the GGA between Ref. \cite{PhysRevLett.126.127601} and this
work can be largely attributed to the differences in the pseudopotentials.
\begin{table}[t!]
	\begin{ruledtabular}
		\begin{tabular}{c|rrrrrr}
			Atom & \multicolumn{2}{c}{$xx,xx$} & \multicolumn{2}{c}{$xx,yy$} & 
			\multicolumn{2}{c}{$xy,xy$} \\ 
			&\multicolumn{1}{c}{LDA}&\multicolumn{1}{c}{GGA}
			&\multicolumn{1}{c}{LDA}&\multicolumn{1}{c}{GGA}
			&\multicolumn{1}{c}{LDA}&\multicolumn{1}{c}{GGA}\\
			\hline
			Li & 0.95 &0.75 &10.22& 9.99 &0.30 &0.32\\ 
			Os & 62.97 &58.18& 24.77& 22.67 &16.02& 15.90 \\
			O$_1$ &$-$1.08& $-$0.99 &$-$2.80 &$-$2.44 &$-$3.28 &$-$3.37\\
			O$_2$ & $-$1.08 &$-$0.99& 9.84 &10.13& $-$0.63 &$-$0.01 \\
			O$_3$ & 85.00 &71.51& 6.17 &5.07& 1.54 &2.70\\\hline
			$\bar{C}_{\alpha\gamma,\beta\delta}$  &437.42& 364.05&143.67 &128.73&41.57& 44.04 \\
			$\bar{C}^\text{HWRV}_{\alpha\gamma,\beta\delta}$  &437.44 &364.19 &143.31 &128.63 &41.83& 44.16\\ 
			Ref. \cite{PhysRevLett.126.127601}&$\dots$&364.7&$\dots$&129.5 &$\dots$&44.3\\\hline
			$f_{\alpha\gamma,\beta\delta}$&$-$14.10 &$-$14.24  & 49.65 &47.56 & 3.69& 3.33 \\
			Ref. \cite{PhysRevLett.126.127601} &$\dots$&$-$13.8&$\dots$&49.3&$\dots$&3.3
		\end{tabular}
		\caption{Linearly independent components of the flexoelectric 
			force-response tensor (in eV), the clamped-ion elastic tensor (in GPa) and
			the flexocoupling tensor (in eV) for 
			cubic LiOsO$_3$. Values are obtained either with the 
			Perdew-Wang parametrization of the LDA or with the PBE
			parametrization of the GGA.}
		\label{table:cubic_20k}
	\end{ruledtabular}
\end{table}
\section{Conclusions}\label{Sec_conclusions}
By combining the virtues of ensemble density-functional theory~\cite{PhysRevLett.79.1337} and density-functional
perturbation theory~\cite{RevModPhys.73.515}, we have established
a general and powerful first-principles approach for higher-order derivatives of
the total energy in metals. 
We have focused our numerical tests on the calculation of the flexoelectric
coupling coefficients, where our formalism offers drastic improvements, 
both in terms of accuracy and computational efficiency, compared to
earlier approaches.
Thereby, our method will greatly facilitate the first-principles-based 
modeling of polar and ferroelectric metals, which are currently under 
intense scrutiny within the research community.
This work also opens numerous exciting avenues for future work; we shall 
outline some of them in the following.

First, the advantages of the approach presented here
can be immediately extended to other adiabatic spatial dispersion 
properties via minor modifications to our formulas.
For example, by combining the phonon perturbation with a 
scalar potential one in Eq.~(\ref{Eq_E2_1q}), our method would yield the 
``adiabatic Born effective charges'' as defined in 
Ref. \cite{marchese2024born,PhysRevLett.129.185902}. The 
present approach works directly at the $\Gamma$ point, and hence  
avoids the need for cumbersome numerical fits.
On the other hand, by targeting the adiabatic response to a static
(but spatially nonuniform) vector potential field, the present 
theory could be used to generalize the theory of orbital magnetic 
susceptibility of Ref. \cite{PhysRevB.84.064445} to metals.

Second, note that the scopes of our work go well 
beyond the specifics of long-wavelength expansions.
Our main conceptual achievement consists in generalizing the ``$2n+1$'' 
theorem~\cite{RevModPhys.73.515}, one of the mainstays of DFPT in insulators, to metallic systems.
This result opens exciting opportunities for calculating not only
spatial dispersion effects, but also \emph{nonlinear response} properties
in metals, with comparable advantages at the formal and practical level.
The study of nonlinear optics appears as a particularly attractive topic in this context,
although its inherent dynamical nature would require generalizing the 
formalism presented here to the nonadiabatic regime. We regard this as
a promising avenue for future developments of our method.
\begin{acknowledgments}
	We acknowledge support from Ministerio de Ciencia
	e Innovaci\'on (MICINN-Spain) through
	Grant No. PID2019-108573GB-C22;
	from Severo Ochoa FUNFUTURE center of excellence (CEX2019-000917-S) and
	from Generalitat de Catalunya (Grant No. 2021 SGR 01519).
\end{acknowledgments}
\appendix
\section{Treatment of the macroscopic electrostatic term in the $\mathbf{q}\rightarrow\mathbf{0}$ limit}\label{Appendix_electrostatics}

\subsection{Long-wave limit of external potentials}
We shall start by reviewing the long-wavelength behavior of the external potentials, 
highlighting the differences between insulators and metals.
\subsubsection{Insulators}
The external potential at first-order in response to an atomic displacement perturbation 
is usually expressed as a sum of a local plus a separable part \cite{PhysRevB.55.10337,PhysRevX.9.021050}. For our scopes, the latter can be omitted,
as no divergences associated to the separable part are present in the long-wave limit. The macroscopic
component ($\mathbf{G=0}$) of the local part is given by \cite{PhysRevB.55.10337}
\begin{equation}\label{Eq_local}
V_\mathbf{q}^{\text{loc},\tau_{\kappa\alpha}}(\mathbf{G=0})\sim
-i\frac{q_\alpha}{\Omega}\left(
-\frac{4\pi Z_\kappa}{q^2}+\frac{F''_\kappa}{2}
\right),
\end{equation}
where $F_\kappa''$ is the second derivative in $q$ of $F_\kappa(q)=q^2v_\kappa^\text{loc}(q)$, with
$v_\kappa^\text{loc}(q)\sim -4\pi Z_\kappa/q^2$,
and
$Z_\kappa$ is the bare nuclear pseudo-charge.
The Hartree potential, on the other hand, is given by
\begin{equation}
V_\mathbf{q}^{\text{H},\tau_{\kappa\alpha}}=\frac{4\pi}{q^2}
\rho_{\tau_{\kappa\alpha}}^\mathbf{q},
\end{equation}
where the lower terms in the Taylor expansion of the first-order electron density 
(in powers of $\mathbf{q}$) are given by
\begin{equation}
\rho^\mathbf{q}_{\tau_{\kappa\alpha}}\sim -iq_\gamma \rho^{(1,\gamma)}_{\kappa\alpha}
-\frac{q_\gamma q_\delta}{2}\rho^{(2,\gamma\delta)}_{\kappa\alpha}+\dots
\end{equation}
The sum of the local and Hartree potentials then reads as (we omit terms that vanish in the
$\mathbf{q}\rightarrow\mathbf{0}$ limit)
\begin{equation}\label{Eq_loc_H}
V_\mathbf{q}^{\text{loc}+\text{{H}},\tau_{\kappa\alpha}}\simeq 
\frac{4\pi}{\Omega}\left(
\frac{iq_\gamma Z^{(\gamma)}_{\kappa\alpha}}{q^2}+\frac{q_\gamma q_\delta}{2q^2}
Q^{(\gamma\delta)}_{\kappa\alpha}
\right),
\end{equation}
where the tensors $Z^{(\gamma)}_{\kappa\alpha}$ and $Q^{(\gamma\delta)}_{\kappa\alpha}$ are,
respectively, the \textit{screened} (short-circuit electrical boundary conditions are assumed) ``Born effective charges" and ``dynamical quadrupoles",
\begin{equation}
\begin{split}
Z^{(\gamma)}_{\kappa\alpha}&=Z_\kappa\delta_{\alpha\gamma}-\Omega\rho^{(1,\gamma)}_{\kappa\alpha},\\
Q^{(\gamma\delta)}_{\kappa\alpha}&=-\Omega\rho^{(2,\gamma\delta)}_{\kappa\alpha}.
\end{split}
\end{equation}
As a consequence, as long as the ``Born effective charges" do not vanish,
in an insulator the potential given by Eq. (\ref{Eq_loc_H}) diverges
as $\mathcal{O}(q^{-1})$, corresponding to the well-known Fr\"olich term in 
the scattering potential.
\subsubsection{Metals}\label{Sec_App_metals}
As we already declared in Sec. \ref{Sec_Fermi_shifts}, in metals the potential should
be an analytic function of the wave vector $\mathbf{q}$, which implies that the divergences that
we have encountered in Eq. (\ref{Eq_loc_H}) should disappear. This involves
\begin{equation}
Z^{(\gamma)}_{\kappa\alpha}=0,\quad 
\Omega\rho^{(1,\gamma)}_{\kappa\alpha}=\delta_{\alpha\gamma}Z_\kappa.
\end{equation} 
In addition, the quadrupoles must be isotropic,
\begin{equation}
Q^{(\gamma\delta)}_{\kappa\alpha}=\delta_{\delta\gamma}Q_{\kappa\alpha}.
\end{equation}
We reach the conclusion that in a metal, the first-order electron density in response to
an atomic displacement  acquires the following form,
\begin{equation}
\Omega\rho^\mathbf{q}_{\kappa\alpha}\sim -iq_\alpha Z_\kappa +\frac{q^2}{2}Q_{\kappa\alpha}.
\end{equation}
When summing the local and the Hartree potential terms, the 
divergencies cancel out and, at leading order in $\mathbf{q}$, the scattering potential
tends to a direction-independent constant,
\begin{equation}
V_\mathbf{q}^{\text{loc+H},\tau_{\kappa\alpha}}\sim \frac{2\pi}{\Omega}Q_{\kappa\alpha},
\end{equation}
which is uniquely determined by the charge neutrality of the unit cell and corresponds
to the Fermi level shifts defined in Sec. \ref{Sec_Fermi_shifts},
\begin{equation}
\frac{2\pi}{\Omega}Q_{\kappa\alpha}=-\mu^{\tau_{\kappa\alpha}}.
\end{equation}
\subsection{The macroscopic electrostatic energy}
\subsubsection{Phonons}
We want to take the first $\mathbf{q}$ derivative of the three terms that contribute
to the macroscopic electrostatic energy in metals, within the framework of variational spatial dispersion
theory. To this end, we shall 
write down the finite $\mathbf{q}$ expressions and we shall take the $\mathbf{q}\rightarrow\mathbf{0}$ limit once the divergencies coming from all the three terms have been properly treated.

The first contribution to the macroscopic electrostatics comes from the ion-ion Ewald (Ew) term,
\begin{equation}
\begin{split}
E_{\text{Ew},\mathbf{q}}^{\tau_{\kappa\alpha}\tau_{\kappa'\beta}}
(\mathbf{G=0})&=Z_\kappa Z_{\kappa'}\frac{4\pi}{\Omega}\frac{q_\alpha q_\beta}{q^2}
e^{-\frac{q^2}{4\Lambda^2}}\\
&\simeq Z_\kappa Z_{\kappa'} \frac{4\pi}{\Omega}
\left(
\frac{q_\alpha q_\beta}{q^2}-\frac{q_\alpha q_\beta}{4\Lambda^2}+\dots
\right),
\end{split}
\end{equation}
where the dots stand for an analytic sum of higher-order terms containing even powers of $\mathbf{q}$.
Its partial derivative with respect to the wave vector $q_\gamma$ is
\begin{widetext}
	\begin{equation}\label{Eq_Ew_deriv}
	E_{\text{Ew},\gamma}^{\tau_{\kappa\alpha}\tau_{\kappa'\beta}}\simeq
	\left[
	\frac{4\pi}{\Omega}\frac{Z_\kappa Z_{\kappa'}}{q^2}\left(
	\delta_{\alpha\gamma}q_\beta + \delta_{\beta\gamma}q_\alpha
	\right)
	-\frac{8\pi}{\Omega}Z_{\kappa}Z_{\kappa'}
	\frac{q_\alpha q_\beta q_\gamma}{q^4}
	\right]_\mathbf{q=0}.
	\end{equation}
\end{widetext}
In Eq. (\ref{Eq_Ew_deriv}) and in the following derivations, we omit terms that vanish in the 
$\mathbf{q=0}$ limit, and we also exclude the $\mathbf{G=0}$ label in order to keep the notation as simple as possible. 

The second contribution comes from the second line of 
Eq. (\ref{Eq_E2_q}), which we shall refer to as
the ``elst" term, and can be equivalently written in reciprocal space as
\begin{equation}\label{Eq_elst_q}
E_{\text{elst},\mathbf{q}}^{\tau_{\kappa\alpha}\tau_{\kappa'\beta}}=
\Omega (\rho^\mathbf{q}_{\kappa\alpha})^* K_\mathbf{q}\rho^\mathbf{q}_{\kappa'\beta},
\end{equation}
where $K_\mathbf{q}=4\pi/q^2$ is the Coulomb kernel. The partial derivative of
Eq. (\ref{Eq_elst_q}) with respect to $q_\gamma$, within the context of our variational 
spatial dispersion theory, i.e., excluding the partial $\mathbf{q}$ derivatives of the 
first-order electron densities, is given by
\begin{widetext}
	\begin{equation}
	\begin{split}
	E_{\text{elst},\gamma}^{\tau_{\kappa\alpha}\tau_{\kappa'\beta}}&=
	\left[\Omega (\rho^\mathbf{q}_{\tau_{\kappa\alpha}})^*K_\gamma \rho^\mathbf{q}_{\kappa'\beta}
	\right]_\mathbf{q=0}\\
	&\simeq\frac{1}{\Omega}
	\left(iq_\alpha Z_\kappa +\frac{q^2}{2}Q_{\kappa\alpha}\right)
	\left(-\frac{8\pi q_\gamma}{q^4}\right)
	\left(-iq_\beta Z_{\kappa'} +\frac{q^2}{2}Q_{\kappa'\beta}\right)\bigg|_\mathbf{q=0}\\
	&\simeq \left[-\frac{8\pi}{\Omega}Z_\kappa Z_{\kappa'}\frac{q_\alpha q_\beta q_\gamma}{q^4}
	+2\left( 
	i\delta_{\alpha\gamma}Z_\kappa \mu^{\tau_{\kappa'\beta}}
	-i\delta_{\beta\gamma}\mu^{\tau_{\kappa\alpha}}Z_{\kappa'}
	\right)\right]_\mathbf{q=0}.
	\end{split}
	\end{equation}
\end{widetext}

In order to compute the third and last contribution to the macroscopic electrostatic energy, which comes from the local (loc) potentials in the
first line of the occupation term, Eq. (\ref{Eq_occ_longwave}), we need the 
first $\mathbf{q}$ derivative of the local part of the psuedopotential given 
in Eq. (\ref{Eq_local}),
\begin{equation}
V_\gamma^{\text{loc},\tau_{\kappa\alpha}}\simeq \frac{4\pi i}{\Omega}
\left[
Z_\kappa\frac{\delta_{\alpha\gamma}}{q^2}-2Z_\kappa\frac{q_\alpha q_\gamma}{q^4}
\right]_\mathbf{q=0}.
\end{equation}
This leads to
\begin{widetext}
	\begin{equation}
	\begin{split}
	E_{\text{loc},\gamma}^{\tau_{\kappa\alpha}\tau_{\kappa\beta}}&=
	\left[
	\Omega (\rho^\mathbf{q}_{\kappa\alpha})^*V_\gamma^{\text{loc},\tau_{\kappa'\beta}}+
	\Omega(V_\gamma^{\text{loc},\tau_{\kappa\alpha}})^*\rho^\mathbf{q}_{\kappa'\beta}
	\right]_\mathbf{q=0}\\
	&\simeq \left[
	\left(iq_\alpha Z_\kappa +\frac{q^2}{2}Q_{\kappa\alpha}\right)\frac{4\pi i}{\Omega}
	\left(Z_{\kappa'}\frac{\delta_{\beta\gamma}}{q^2}
	-2Z_{\kappa'}\frac{q_\beta q_\gamma}{q^4}\right)
	-\frac{4\pi i}{\Omega}
	\left(
	Z_\kappa\frac{\delta_{\alpha\gamma}}{q^2}-2Z_\kappa\frac{q_\alpha q_\gamma}{q^4}\right)
	\left(-iq_\beta Z_{\kappa'}+\frac{q^2}{2}Q_{\kappa'\beta}\right)
	\right]_\mathbf{q=0}\\
	&\simeq \left[-\frac{4\pi}{\Omega}\frac{Z_\kappa Z_{\kappa'}}{q^2}\left(
	\delta_{\alpha\gamma}q_\beta + \delta_{\beta\gamma}q_\alpha\right)
	+\frac{16\pi}{\Omega}Z_\kappa Z_{\kappa'}\frac{q_\alpha q_\beta q_\gamma}{q^4}
	+i\delta_{\beta\gamma}\mu^{\tau_{\kappa\alpha}}Z_{\kappa'}
	-i\delta_{\alpha\gamma}Z_\kappa\mu^{\tau_{\kappa'\beta}}\right]_\mathbf{q=0}.
	\end{split}
	\end{equation}
\end{widetext}
When the three contributions are treated together, all the divergences cancel out, and we are left
with the following final result,
\begin{equation}
\begin{split}
E_{\text{mac},\gamma}^{\tau_{\kappa\alpha}\tau_{\kappa'\beta}}&=
E_{\text{Ew},\gamma}^{\tau_{\kappa\alpha}\tau_{\kappa'\beta}}
+E_{\text{elst},\gamma}^{\tau_{\kappa\alpha}\tau_{\kappa'\beta}}
+E_{\text{loc},\gamma}^{\tau_{\kappa\alpha}\tau_{\kappa'\beta}}\\
&\simeq i\delta_{\alpha\gamma}Z_\kappa\mu^{\tau_{\kappa'\beta}}
-i\delta_{\beta\gamma}\mu^{\tau_{\kappa\alpha}}Z_{\kappa'},
\end{split}
\end{equation}
which is Eq. (\ref{Eq_correction_Fermi_shift_deriv}) of the main text.
\subsubsection{Strain gradients}
It is also interesting for our scopes to analyze the contributions coming from the macroscopic
electrostatics at second-order in $\mathbf{q}$.  By following the same strategy as in 
Sec. \ref{Sec_strain_gradients}, second derivatives in $\mathbf{q}$ can be avoided by
treating the acoustic phonon perturbation in the comoving frame as a metric-wave perturbation.
The final result for the macroscopic electrostatic contribution reads as
\begin{equation}
\begin{split}
E_{\text{mac},\gamma}^{\tau_{\kappa\alpha}\eta_{\beta\delta}}&=
E_{\text{elst},\gamma}^{\tau_{\kappa\alpha}\eta_{\beta\delta}}+
E_{\text{loc},\gamma}^{\tau_{\kappa\alpha}\eta_{\beta\delta}}\\
&\simeq -\delta_{\alpha\gamma}Z_\kappa \mu^{\eta_{\beta\delta}},
\end{split}
\end{equation}
where $\mu^{\eta_{\beta\delta}}$ is the Fermi level shift produced by a uniform strain.


\input{ensemble_DFPT.bbl}

\end{document}

%% file: ensemble_DFPT.bbl
%